\documentclass[11pt]{article}
\usepackage{jheppub}
\usepackage[utf8]{inputenc}
\usepackage{amsfonts}
\usepackage{amsmath}
\usepackage{amssymb,amsthm}
\usepackage{wasysym}
\usepackage{mathtools}
\usepackage{array}
\usepackage{comment}
\usepackage{hyperref}
\newtheorem{property}{Property}
\usepackage{booktabs}
\usepackage{graphicx}
\usepackage{float}
\usepackage{longtable}
\usepackage{adjustbox}
\usepackage{multirow}
\usepackage{pdflscape}
\usepackage{rotating}

\usepackage{soul}

\def\blue#1{{\color{blue}{#1}}}

\newcommand*\modulo[1]{\quad(\mathrm{mod}\, #1)}
\newcommand*\diff{\mathop{}\!\mathrm{d}}
\theoremstyle{definition}
\newtheorem{example}{Example}[section]
\usepackage{soul}

\title{Allowed Coulomb branch scaling dimensions \\ of four-dimensional $\mathcal{N}=2$ SCFTs} 
\author[a]{Philip C. Argyres,}
\author[b,c,d]{Sergio Cecotti,}
\author[e,f,g]{Michele Del Zotto,}
\author[h,i]{Mario Martone,}
\author[j]{Robert Moscrop,}
\author[h]{and Ben Smith}

\affiliation[a]{University of Cincinnati,
Physics Department, Cincinnati OH 45221}
\affiliation[b]{Beijing Institute of Mathematical Sciences and Applications (BIMSA), Huaibei Town, Huairou District, Beijing 101408, China}
\affiliation[c]{Qiuzhen College, Tsinghua University, Haidian District, Beijing, China}
\affiliation[d]{SISSA, via Bonomea 265, I-34100 Trieste, Italy} 
\affiliation[e]{Mathematics Institute, Uppsala University,
Box 480, SE-75106 Uppsala, Sweden}
\affiliation[f]{Department of Physics and Astronomy, Uppsala University,
Box 516, SE-75120 Uppsala, Sweden}
\affiliation[g]{Center for Geometry and Physics, Uppsala University,
Box 480, SE-75106 Uppsala, Sweden}
\affiliation[h]{Dept. of Mathematics, King’s College London, The Strand, London WC2R 2LS, UK}
\affiliation[i]{EliseAI, New York, NY 10016, USA}
\affiliation[j]{Center of Mathematical Sciences and Applications, Harvard University, 20 Garden Street, MA  02138,  USA}

\emailAdd{philip.argyres@gmail.com}
\emailAdd{cecotti@bimsa.cn}
\emailAdd{michele.delzotto@math.uu.se}
\emailAdd{mario.martone@kcl.ac.uk}
\emailAdd{robert@cmsa.fas.harvard.edu}
\emailAdd{benjamin.g.smith@kcl.ac.uk}

\abstract{A basic datum of a rank-$r$ $\mathcal{N}{=}2$ superconformal field theory (SCFT) is the $r$-tuple of its Coulomb branch scaling dimensions, i.e., the scaling dimensions of a set of special protected scalar operators whose vevs generate the coordinate ring of the Coulomb branch of the theory. 
It is well known that when the coordinate ring is freely generated these scaling dimensions can only take values in a small set of rational numbers.
But there are further constraints on which $r$-tuples of these numbers can appear. 
The main aim of this work is to clarify what these are. 
Along the way we also compute explicitly the $r$-tuples of allowed scaling dimensions for theories of ranks $r=2,3,4$.}

\begin{document}
\maketitle

\section{Introduction and summary}

The space of unitary four-dimensional $\mathcal{N}{=}2$ superconformal field theories (SCFTs) of non-zero rank appears to be remarkably constrained by the possible Coulomb branch (CB) geometries of their moduli spaces, characterized and constrained by their rigid special K\"ahler (SK) structures and scaling symmetries. 
When a rank-$r$ (i.e., $r$-complex dimensional) CB is free of complex singularities, its coordinate ring is freely generated by the vacuum expectation values (vevs) $\{u_i, \ i=1, \dots, r\}$ of a special set of operators, the \emph{Coulomb branch operators}, with definite scaling dimensions $\{\Delta_i\}$.
The problem of determining which $r$-tuples of CB dimensions are allowed is the subject of this paper.

It has been shown that the scaling dimensions for a rank-$r$ CB are restricted to take values in the finite set of rational numbers \cite{Caorsi:2018zsq, Argyres:2018urp}
\begin{equation}\label{Set}
    \Delta_j \in \left\{ \frac{n}{m} \ \bigg| \ 0 < m \leq n, \ \varphi(n) \leq 2r, \ \gcd(m,n) = 1 \right\},
\end{equation}
where $\varphi$ is the Euler totient function.
(We re-derive this result in section \ref{sec:constrain}.)
However, not all sets of $r$-many such values give rise to consistent CB geometries.
In fact, the allowed $r$-tuples of dimensions represent a significantly smaller subset, thus making the CB scaling dimension $r$-tuple an even handier tool to preliminarily determine the consistency of a given candidate rank $r$ $\mathcal{N}{=}2$ SCFT, heavily restricting the possible theories in question.
For example, setting $r=2$ in \eqref{Set} gives the 23 allowed dimensions
\begin{align}\label{eq:rank2}
\Delta_j^{r=2}\in\left\{\frac{12}{11},\frac{10}{9},\frac{8}{7},\frac{6}{5},\frac{5}{4},\frac{4}3,\frac{10}{7},\frac{3}{2},\frac{8}{5},\frac{5}{3},\frac{12}{7},2,\frac{12}{5},\frac{5}{2},\frac{8}{3},3,\frac{10}3,4,5,6,8,10,12\right\} .
\end{align}
Thus the na\"ive expectation is that there should be 276 distinct allowed pairs while a closer analysis cuts down this number to 63.%
\footnote{This number differs from the 79 pairs stated in \cite{Kaidi:2022sng}. 
This is because that reference did not account for the \emph{non-reflexivity} of the genuine scaling dimension conditions explained here in section \ref{sec:conds} below equation \eqref{eq:delts}.} 
See, e.g., \cite{Cecotti:2023ksl} for further examples. 
As such, mapping out the consistent set of scaling dimensions at a given rank provides invaluable information for the ongoing classification of $\mathcal{N}{=}2$ SCFTs.

We present here a careful and systematic construction of the full set of algebraic relations which CB operator scaling dimensions have to satisfy, from which the allowed $r$-tuples follow.
The algebraic relations are a consequence of the EM duality monodromies and the structure of singular loci encoded in the special geometry of scale-invariant CBs. In particular, this clarifies previous works \cite{Caorsi:2018zsq, Kaidi:2022sng, Argyres:2022lah} by some of the authors which contained partial or incorrect characterizations of the allowed $r$-tuples.

It is worth reminding the reader that the correspondence between $r$-tuples of CB scaling dimensions and $\mathcal{N}=2$ SCFTs is neither injective nor surjective. 
In fact, there are many known cases where several inequivalent $\mathcal{N}=2$ SCFTs share the same $r$-tuple of scaling dimensions (see, for example, \cite{Martone:2021ixp}). 
Similarly, there is no guarantee that all allowed $r$-tuples are realized. 
While there are $\mathcal{N}=2$ SCFTs realizing all allowed rank-1 scaling dimensions \cite{Argyres:2015ffa, Argyres:2015gha, Argyres:2016xmc}, already at rank-2 there are several couples $(\Delta_1,\Delta_2)$ which have no known realisation (so far) as scaling dimensions of CB operators of a known SCFT.

This paper is organized as follows.
In section \ref{sec2}, we review the most relevant aspects of scale-invariant CB geometries with no complex singularities, focusing on their scaling and special Kahler stratifications.
This review is by no means self-contained, but is meant to bring together and highlight results from \cite{Argyres:2020wmq, Cecotti:2023ksl}.
Next, in section \ref{sec:constrain}, we delve into the way the two stratifications interact to constrain the allowed $r$-tuples.
The resulting constraints on allowed $r$-tuples are summarized around equation \eqref{eq:delts}. 
Finally, section \ref{sec4} presents an algorithm for calculating all physically admitted sets of CB scaling dimensions. 
Appendix \ref{app:tables} tabulates the allowed scaling dimensions at rank-2, rank-3, and rank-4 which follow from this algorithm.
For each rank we only tabulate the genuinely new tuples of scaling dimensions \cite{Martone:2021ixp,Argyres:2022lah,Argyres:2022puv,Argyres:2022fwy}. 
This is a subset of allowed scaling dimensions but it is the essential datum that allows to compute the entire set. 
The precise definition of this subset can be found below in section \ref{gen sec}.

\section{A review of Coulomb branch geometry}\label{sec2}

Other reviews of CB geometry can be found in \cite{Alvarez-Gaume:1996ohl, Lerche:1996xu, Freed:1997dp, Martone:2020hvy}.

\subsection{Scale invariant Coulomb branches without complex singularities}

Any rank-$r$ $4$d $\mathcal{N}{=}2$ SCFT possesses a space of gauge inequivelant vacua known as its moduli space. The moduli space has several branches, some of which special, that can be characterized by the pattern of breaking of the $\mathrm{U}(2)$ superconformal $R$-symmetry. 
In particular, the CB, denoted $\mathcal{C}$, is characterized by preserving the $\mathrm{SU}(2)_R$ symmetry and spontaneously breaking the $\mathrm{U}(1)_R$ symmetry.
It is an $r$-complex-dimensional space with a rigid special K\"ahler (SK) geometry with metric non-analyticities on a subspace $\mathcal{V} \subset \mathcal{C}$ \cite{Freed:1997dp, Seiberg:1994aj, Seiberg:1994rs}. 
Physically, the vacua where the CB metric is non-analytic  are vacua for which the theory has extra massless charged states. 

If we assume that the CB has no complex structure singularities --- an assumption we will be making throughout --- it is always possible to find a set of $r$ complex coordinates, $u:=(u_{1}, \ldots, u_{r})$, which are globally defined on $\mathcal{C}$ and which we will take as coordinates on this space. 
These can also be identified with the vevs of the CB operators --- scalar superconformal primaries which are singlets of $\mathrm{SU}(2)_R$ and which satisfy the shortening condition that their $\mathrm{U}(1)_R$ charge is proportional to their scaling dimension --- which generate the CB chiral ring \cite{Beem:2014zpa}. 
The absence of complex singularities is translated algebraically by the requirement that the CB chiral ring is freely generated \cite{Argyres:2017tmj}. 
If it is not the case, we have relations between the vevs of the generating set of CB operators and a unique set of CB scaling dimensions becomes harder to define. 
Herein we will only consider theories with a freely generated chiral ring.

The CB inherits additional structure from the parent SCFT. 
The scale symmetry from the conformal symmetry group and the $\mathrm{U}(1)_{R}$ symmetry of the superconformal group act non-trivially on the CB. 
We define the action of dilatations on CB operators as $\Phi_{i}(x) \mapsto \lambda^{\Delta_{i}} \Phi_{i}(\lambda x)$, for $\lambda$ a positive real number. 
The weights, $\Delta_j$, of operators under this action are called their scaling dimensions, and are constrained to be greater than or equal to 1 by unitarity.
The scaling dimension is proportional to the $\mathrm{U}(1)_{R}$ charge for CB operators, so they combine to give a complex action on the CB.
In particular, under the assumption of freely generated CB chiral ring this action is generated by the holomorphic vector field
\begin{align}\label{C*cE}
    \mathcal{E} = \sum_j \Delta_j u_j \frac{\partial}{\partial u_j}
\end{align}
where $u_j \doteq \langle\Phi_j\rangle$ are the vevs, assumed to be global complex coordinates on the CB.
This exponentiates to a $\mathbb{C}^\times$ action on $\mathcal{C}$ given by
\begin{align}\label{C*action}
    \mathbb{C}^\times:  u &\mapsto \exp(z\mathcal{E}) \cdot u \doteq \left( e^{z \Delta_{1}}u_{1}, \ldots, e^{z\Delta_{r}}u_{r} \right),
\end{align}
for all $z \in \mathbb{C}$.

The \emph{$r$-tuple of CB scaling dimensions}, $(\Delta_{1}, \ldots, \Delta_{r})$, is the main object of study of this paper.
Although the $(u_i)$ coordinates are not uniquely defined by \eqref{C*action} --- linear redefinitions of $u_i$'s with the same dimension are possible --- the $r$-tuple of scaling dimensions (including multiplicities) is uniquely defined.

Associated to this $\mathbb{C}^\times$ action is a stratification of the CB by its $(u_i)$ coordinate hyperplanes (see \cite{Argyres:2020wmq} for an in depth discussion).
In particular, for each subset $(i_1,\ldots,i_\ell) \subset (1,\ldots,r)$, define a dimension-$\ell$ hyperplane in $\mathcal{C} \simeq \mathbb{C}^r$ by
\begin{align}\label{C*strata}
    \mathcal{I}_{i_1,\ldots,i_\ell} &\doteq \{u\in\mathbb{C}^r\,|\, u_j = 0, \ \forall  j \notin (i_1,\ldots,i_\ell) \} .
\end{align}
We will call the subset of scaling dimensions $(\Delta_{i_1}, \ldots \Delta_{i_\ell})$ those \emph{associated} to $\mathcal{I}_{i_1,\ldots,i_\ell}$.
We will be particularly interested in the 1-dimensional \emph{$u_i$-coordinate axis} $\mathcal{I}_i$ with associated dimension $\Delta_i$.

The interplay of this $\mathbb{C}^\times$ stratification with the stratification of the CB generated by its metric non-analyticities will be key to our investigation. 
This latter stratification --- the \emph{SK stratification} --- is quite constrained \cite{Argyres:2020wmq, Cecotti:2023ksl} by the SK structure of the CB and the physical interpretation of the CB non-analyticities, as we now review.
We start by reviewing the SK structure of the CB away from its metric non-analyticities.

\subsection{SK geometry of the CB and its associated algebraically integrable system}

The states in the theory at a generic point of the CB are specified by their charges under the $\mathrm{U}(1)^{r}$ gauge symmetry. 
We will collectively call these $\mathbf{p}$. Dirac quantisation restricts $\mathbf{p}$ to lie in a lattice $\Lambda\cong\mathbb{Z}^{2r}$ equipped with an antisymmetric integer pairing called the Dirac pairing, $J$. 
Each point on the charge lattice has a pair of electric and magnetic charges for each $\mathrm{U}(1)$ gauge factor. 
The group that preserves $\Lambda$ and a given Dirac pairing $J$, the electric-magnetic (EM) duality group, is $\mathrm{Sp}_J(2r, \mathbb{Z})$. 
If the pairing is principal, this is the usual symplectic group $\mathrm{Sp}(2r, \mathbb{Z})$.%
\footnote{Theories with non-principal pairings are so-called \emph{relative theories} \cite{Freed:2012bs, Aharony:2013hda}. 
See, e.g., \cite{Gaiotto:2010be, DelZotto:2022ras, Argyres:2022kon, Closset:2023pmc} for more on 4d relative QFTs.} 
$\mathrm{Sp}_J(2r,\mathbb{Z})$ duality transformations leave the physics of the $\mathrm{U}(1)^{r}$ theory invariant. 

Vevs of the scalar superpartners of the $\mathrm{U}(1)^r$ field strengths (in an ``electric'' EM duality frame) are special coordinates, $a^i(u)$, on the CB.
In a magnetic duality frame, they are dual special coordinates, $a_i^D(u)$.
These transform under the $\mathrm{Sp}_J(2r, \mathbb{Z})$ EM duality group as holomorphic vector-valued functions on the CB,
\begin{equation}\label{SC}
    \sigma(u) \doteq \left(\begin{smallmatrix} \mathbf{a}^{D}(u)\\ \mathbf{a}(u) \end{smallmatrix}\right).
\end{equation}
Unbroken $\mathcal{N}{=}2$ supersymmetry implies that the matrix of low-energy $\mathrm{U}(1)^{r}$ complex gauge couplings is given in terms of the special coordinates by
\begin{equation}\label{tauij def}
    \tau_{ij}(u) = \frac{\partial a_{i}^D}{\partial a^{j}}.
\end{equation}
The K\"ahler metric components on the CB (i.e., the kinetic terms of the scalars) with respect to the $a^j$ special coordinates are given by $g_{i j} = \text{Im}(\tau_{ij})$.
This identification together with unitarity imply that the coupling matrix $\tau_{ij}$ is symmetric and has positive definite imaginary part.
These structures and properties define the special K\"ahler (SK) structure of the CB.

The SK structure can be succinctly described in terms of the \emph{special section} $\sigma$ of an $\mathrm{Sp}_J(2r, \mathbb{Z})$ vector bundle over $\mathcal{C}$ whose fiber, $V^{*}$, is the linear dual of the complexification of the charge lattice \cite{Argyres:2018zay}.
The complexification of the charge lattice inherits both the Dirac pairing and the action of the EM duality group.
Then the CB K\"ahler potential is $K= i J(\sigma, \bar{\sigma})$, with \eqref{tauij def} and the symmetry of $\tau_{ij}$ is ensured by the condition $J(\partial_{i} \sigma, \partial_{j} \sigma) = 0$.
Positive-definiteness of the K\"ahler metric ensures positivity of $\text{Im}(\tau_{ij})$.
Note that the special coordinates are linearly related to the central charge of the $\mathcal{N}=2$ SUSY algebra via the BPS mass formula. As such, the weight of $\sigma(u)$ under the $\mathbb{C}^\times$ action is $\Delta_{\sigma} = 1$.
Note also that $\sigma$ does not diverge anywhere in $\mathcal{C}$, as this would give a sub-sector of the theory that is decoupled at all scales. 

The seminal works of \cite{Seiberg:1994aj, Seiberg:1994rs, Donagi:1995am, Donagi:1995cf} associate an \emph{algebraically integrable system} to the non-singular locus of the CB; that is, a fibration of polarized abelian varieties $X_u$ over the CB $\mathcal{C}$ endowed with a holomorphic symplectic two-form $\omega$ that vanishes when restricted to the fibers. 
The latter property can be summarized as saying that the fibration is Lagrangian with respect to $\omega$. 
Let us briefly discuss this notion, before using it to constrain the possible sets of CB scaling dimensions in section \ref{sec:constrain}. 
For more detailed accounts of these notions, we refer the reader to \cite{Donagi:1995am, lange2013complex} and to appendix A of \cite{Argyres:2024hdn}.

An abelian variety is a complex torus $A=\mathbb{C}^r/\Lambda$ that is also a projective variety. 
As such, in order to fully specify an abelian variety, we must also state how to embed the complex torus $A$ into $\mathbb{P}^{n}$ for some  $n\in\mathbb{N}$. 
This can be achieved by equipping $A$ with an integral non-degenerate skew pairing $J$ on $\Lambda$.%
\footnote{Geometrically, $J$ corresponds to the first Chern class of an ample line bundle $\mathcal{L}$ on $A$. As $\mathcal{L}$ is ample, the sections of an appropriate power of it defines a closed immersion of $A$ into $\mathbb{P}^n$ for some $n$.} 
Such a $J$ is called a \emph{polarization} on $A$ and is said to be \emph{principal} if $\det J=1$. 
In the context of $\mathcal{N}{=}2$ SQFTs, the electromagnetic charge lattice $\Lambda_u$, IR effective couplings $\tau_{ij}(u)$, and the Dirac pairing $J$ provide the datum of a polarized abelian variety associated to a point $u\in\mathcal{C}$. 

Concretely, define $X_u= \mathbb{C}^r/\Lambda_u$, and let $\{\eta_i\}$ be a basis of $H^{(1,0)}(X_u)\cong \mathbb{C}^r$ and $\{\alpha^i, \beta_j\}$ be a basis of $H_1(X_u) \cong \Lambda_u$.
Define the \emph{period matrix}, $\Pi$, of $X_u$ with respect to these bases as the $(r\times 2r)$-dimensional matrix
\begin{align}\label{per mat}
    \Pi = \big({\textstyle \int_{\boldsymbol{\alpha}} \boldsymbol{\eta}, \int_{\boldsymbol{\beta}} \boldsymbol{\eta}}\big) .
\end{align}
By the Riemann conditions, $X_u$ is an abelian variety if and only if there are bases such that $\Pi = \big(J,\tau\big)$, with $J$ and $\tau_{ij}$ satisfying the SK conditions described above.
In this way the SK structure defines a holomorphic fibration of abelian varieties over the CB.
Note that the non-degeneracy of $J$ and $\text{Im}\,\tau_{ij}$ imply that 
\begin{align}\label{detP}
    \det \left( \begin{smallmatrix} \Pi \\ \Bar{\Pi} \end{smallmatrix} \right) \neq 0,
\end{align}
which is just the statement that $\Lambda_u$ is a full rank lattice in $\mathbb{C}^r$, so $X_u$ is a non-degenerate torus.

Furthermore, if $\omega$ is a holomorphic two-form on the total space of this fibration of abelian varieties over $\mathcal{C}$ with vanishing restriction to the fibers, then its fiber integrals are well-defined.
If we identify these fiber integrals with the CB differentials of the special section,
\begin{gather}\label{eq:int_omega}
    \diff a^i(u) = \int_{\alpha^i} \omega,\quad \diff a^D_i(u) = \int_{\beta_i} \omega,
\end{gather}
then the SK conditions on the special coordinates follow from $\omega$ being closed and non-degenerate on the total space.
This means that $\omega$ is a holomorphic symplectic form with respect to which the fibration is lagrangian.
In this way the algebraically integrable system is defined by the CB SK geometry.

Due to the fact that $\omega$ is a symplectic form, we can view $\omega^{-1}$ as giving a map from the (holomorphic) cotangent space of the CB to the tangent plane of the fiber at $u$,
\begin{equation}\label{eq:omega}
    \omega^{-1} : T^{*}_{u} \mathcal{C} \rightarrow T_\mu X_{u},
\end{equation}
where $\mu$ is a generic point on $X_u$, thus giving an isomorphism of vector spaces. 

\subsection{SK stratification of the CB}

In extended supersymmetry, a non-zero SUSY central charge implies a non-trivial lower (BPS) bound on the masses, $M$, of charged states. 
In the case that we are analysing here, the central charge, for a given vacuum $u \in \mathcal{C}$ and charge  $\mathbf{p} \in \Lambda_u$, is given by \cite{Witten:1978mh}
\begin{equation}
    Z_{\mathbf{p}}(u) = \mathbf{p}^{T} \sigma(u).
\end{equation}
It follows from the BPS bound, $M \geq |Z|$, that states with charge $\mathbf{p}$ can only become massless at zeros of the locally holomorphic function $Z_{\mathbf{p}}(u)$.
As this would result in massless charged states in the effective $\mathrm{U}(1)^{r}$ theory, the IR effective action, written in terms of free vector multiplets, breaks down, as reflected in non-analyticities of the CB metric. 
\textit{We assume all non-analyticities of the CB, $\mathcal{V} \subset \mathcal{C}$, have this form.}\footnote{\ Metric singularities at finite distance in moduli space that are not associated to the presence of extra particle-like massless degrees of freedom (corresponding to a vanishing central charge) would indicate the theory at hand has potentially more exotic massless degrees of freedom, such as a tensionless string.}
Non-analyticities of the metric along $\mathcal{V}$ imply $Z_{\mathbf{p}}$ is non-analytic there for those charges $\mathbf{p}$ corresponding to BPS states in the spectrum.
This, in turn, implies non-analyticities of the special coordinates $\sigma$ of a special form at $\mathcal{V}$.
Requiring the IR effective action to be physically consistent in the vicinity of $\mathcal{V}$ implies \cite{Argyres:2018urp}:
\begin{itemize}
    \item $\mathcal{V}$ is closed in $\mathcal{C}$. If it were not, there would be no consistent physical interpretation of the IR effective action at the boundary points which are not contained within $\mathcal{V}$.
    \item The K\"ahler metric extends over $\mathcal{V}$. It follows all distances on $\mathcal{V}$ are finite and well-defined.
    \item As $Z_{\mathbf{p}} = 0$ on $\mathcal{V}$, we can take $\mathcal{V}$ to be a union over components which vanish for a given value $\mathbf{p}$.
\end{itemize}

Further we assume $\mathcal{V}$ is a complex analytic set in $\mathcal{C}$. 
This is to ensure the non-existence of accumulation points in the complex plane transverse to $\mathcal{V}$. 
This assumption may not be necessary, as it might follow from the local holomorphicity of $\mathcal{V}$ and the fact that there are only a countably infinite number of central charges whose zeros can define $\mathcal{V}$ \cite{Argyres:2018urp}. 
We can conclude from this that $\mathcal{V}$ is a complex co-dimension 1 subset of $\mathcal{C}$, and a generic point in $\mathcal{V}$ is a regular complex hypersurface in $\mathcal{C}$. We further assume that $\mathcal{V}$ is actually algebraic in $\mathcal{C}$.
We can write $\mathcal{V}$ as a union of its co-dimension 1 irreducible components,
\begin{equation}
    \mathcal{V} = \bigcup_{a} \ \mathcal{V}_{a} .
\end{equation}
The non-analyticity of the special section along one $\mathcal{V}_{a}$ component is reflected in there being a non-trivial EM monodromy, $M_a \in \mathrm{Sp}_J(2r, \mathbb{Z})$ around a path $\gamma_a \in \pi_1(\mathcal{C} \setminus \mathcal{V})$ linking only $\mathcal{V}_{a}$.
The characterization of these possible linking monodromies, $M_a$, and their connection to rank-1 ``transverse slice'' SK geometries are described in more detail in \cite{Martone:2020nsy, Argyres:2020wmq, Cecotti:2021ouq}.

But more interesting for our purpose is the fact \cite{Argyres:2020wmq, Cecotti:2023ksl} that the rank-$(r-1)$ $\mathcal{V}_{a}$ subvarieties inherit SK geometries of their own.
These geometries, in turn, have co-dimension-1 singular components (i.e., where their inherited metrics have non-analyticities), and so forth, leading to an \emph{SK stratification} of the CB.
In terms of the $\mathcal{V}_{a}$ subvarieties, (the closure of) a co-dimension $s$ stratum of $\mathcal{C}$ is a connected component of an $s$-fold intersection of co-dimension-1 singular components, $\mathcal{V}_{a_1} \cap \cdots \cap \mathcal{V}_{a_s}$ (meant to include the cases where some of the intersections may be self-intersections).

Before turning to the algebraically integrable system, let us state some further assumptions we make on the special geometry. First, we assume the existence of a section of the abelian fibration over $\mathcal{C}$. This restricts the behavior of the singular fibers, thus allowing a notion of symplectic reduction \cite{Cecotti:2023ksl}.
The existence of such a section is a requirement of a physical CB geometry, and is inherited by the SK stratification.
Also note that for certain ``$I_n$-type'' singular fibers the inherited SK geometry on the stratum is less constrained than those of physical CBs \cite{Argyres:2020wmq, Cecotti:2021ouq}.
(These were called ``irregular geometries'' in \cite{Argyres:2020wmq}.)
As these fibers primarily occur in theories with non-freely generated CBs, their existence does not affect the argument given here, so we will not focus on them.
Secondly, the analysis of \cite{HO2007, HO2009} assumes that the total space of the integrable system is a manifold, which is not obviously a requirement of a physical SK geometry.
The analysis of \cite{Argyres:2020wmq}, though less rigorous, indicates that the SK stratification nevertheless persists.
Finally, in order to continue the SK stratification down in dimensionality, we assume that the singularities of the SK fibrations inherited by the strata are necessarily tame enough to do so. This validity of this assumption is not completely obvious from the physical perspective, but examples seem to support the hypothesis that the SK stratification continues ``down''. It is worth noting that these examples are mostly at low rank, however.

With these assumptions, the SK stratification can be understood in terms of the algebraically integrable system picture of SK geometry as follows \cite{Cecotti:2023ksl}.
According to the analysis of \cite{HO2007, HO2009} the singular fiber, $X_u$, at a regular point of a co-dimension-1 SK stratum, $u \in \mathcal{V}_a$, can be resolved into a set of transversely intersecting components, all of which are fiber bundles with a $\mathbb{P}^1$ fiber over a rank $r{-}1$ polarized abelian variety, $A_u$.
The fibration of $A_u$ over $\mathcal{V}_a$ together with the restriction of the symplectic form is then a rank-$(r{-}1)$ SK geometry in its own right, thus giving the SK stratification.
In particular, the restriction of the inverse symplectic form $\omega^{-1}$ to the resolved singular fiber (i.e., the $\mathbb{P}^1$ fiber bundle over $A_u$) still gives an isomorphism $T_u^*\mathcal{C} \cong T_\mu X_u$ as in \eqref{eq:omega}, though now the tangent space to the (appropriate component of the resolved) singular fiber, $T_\mu X_u$, has direction tangent to its $A_u$ abelian variety base as well as to its $\mathbb{P}^1$ fiber.
In particular, the $(r{-}1)$-dimensional span of the cotangents to the $\mathcal{V}_a$ stratum at $u$ map to the tangent space of the $A_u$ polarized abelian variety in the fiber.\footnote{To be more rigorous, we should think of the bundle of cotangents to the stratum as a quotient of $T^*\mathcal{C}$ by the conormal bundle. This leads to the same conclusions and we refer the reader to section 2 of \cite{Cecotti:2023ksl} for more details.}

Iterating this argument to lower dimensional strata, i.e., by restricting to higher-codimen\-sion subvarieties of the CB with additional metric non-analyticities, we find the following picture of the (resolved) singular fiber there.
On an $\ell$-dimensional stratum, $\mathcal{V}_{\ell;a}$, the resolved singular fiber at a general point $u \in \mathcal{V}_{\ell;a}$ is a collection of intersecting components each of which are fiber bundles over a rank-$\ell$ abelian variety, $A^{\ell;a}_u$.
And the tangent space to one of these fiber bundle components is isomorphic via $\omega^{-1}$ to the cotangent space of the base, with cotangents to the stratum mapped to tangents to $A^{\ell;a}_u$.

As the dilatations and $\mathrm{U}(1)_R$ rotations are symmetries, their $\mathbb{C}^\times$ action on the CB will, in particular, preserve the spectrum of charged states of the theory as well as the low energy effective action on the CB.
This implies that it acts as an isomorphism between the abelian variety fibers of the integrable system and preserves the symplectic form.
That is, the holomorphic vector field $\mathcal{E}$ defined in \eqref{C*cE} extends to one on the total space of the algebraically integrable system such that
\begin{align}\label{cE act fiber}
    e^{z\mathcal{E}} \circ X_u &= X_{e^{z\mathcal{E}}u}, &
    e^{z\mathcal{E}} \circ \omega &= e^z \omega.
\end{align}
This last follows from the fact that weight of $\omega$ under the $\mathbb{C}^\times$ action on $\mathcal{C}$ is $\Delta_\omega=1$ since the fiber periods of $\omega$ are the special coordinates \eqref{eq:int_omega} which have mass dimension 1.
The $\mathbb{C}^\times$ action therefore preserves each co-dimension-1 singular component, $\mathcal{V}_a$, as well as each of the SK strata defined by their intersections.
That is to say, each $\ell$-dimensional SK stratum $\mathcal{V}_{\ell;a}$ is itself a union of orbits of the $\mathbb{C}^\times$ action.

\section{Constraints on tuples from automorphisms of abelian varieties}
\label{sec:constrain}

In this section we will show how to use the $\mathbb{C}^\times$ symmetry action together with the SK stratification of the CB to put strong constraints on the spectrum of possible CB scaling dimensions.
The key results are Properties \ref{fact0} and \ref{fact10}, derived in the next subsection.
These are closely related to earlier results described in \cite{Argyres:2018zay, Caorsi:2018zsq, Argyres:2018urp, Argyres:2020wmq} and Property \ref{fact10} appears as ``Fact 10'' in section 2.12 of \cite{Cecotti:2023ksl}.
Then in subsection \ref{sec:conds} we derive the algebraic constraints on compatible tuples of CB scaling dimensions which follow from these Properties.

\subsection{Genuine rank-\texorpdfstring{$r$}{r} scaling dimensions}
\label{gen sec}

The set of scaling dimensions appearing at a given rank can be ordered by the smallest rank at which they first occur in a CB geometry.
We call a scaling dimension a \emph{genuine rank-$\ell$} (or \emph{$\ell$-genuine}) scaling dimension if it appears in CB geometries with rank $\ell$ and higher.%
\footnote{These are referred to as \emph{new dimensions in rank-$\ell$} in \cite{Cecotti:2023ksl}.}
This is a sensible notion since if a scaling dimension occurs at rank $\ell$, it necessarily occurs at all higher ranks, if only because higher-rank CB geometries can be formed by taking products of lower-rank geometries.

In general, the intersection of the $\mathbb{C}^\times$-strata $\mathcal{I}_{i_1 \ldots i_\ell}$ defined in \eqref{C*strata} with the SK strata $\mathcal{V}_{\ell';a}$ may be very complicated.
But the intersections of SK strata with the 1-dimensional $\mathbb{C}^\times$-strata $\mathcal{I}_i$ --- the $u_i$ coordinate axes --- are simple: since they are a single $\mathbb{C}^\times$ orbit, they either are wholly contained in a given SK stratum or do not intersect it at all.%
\footnote{Technically, for this to be true, we are defining strata as the open sets (manifolds) formed by subtracting from their closure any proper sub-strata.}
A $\mathbb{C}^\times$ stratum $\mathcal{I}_{i_1 \ldots i_\ell}$ comes with \emph{associated dimensions} $(\Delta_{i_1}, \ldots, \Delta_{i_\ell})$, which, recall, are the scaling dimensions of the CB scaling coordinates which are not set to zero on the stratum.

The key properties of the spectrum of CB branch scaling dimensions and the structure of CB singularities which follow from the $\mathbb{C}^\times$ symmetry and the SK stratification are:
\begin{property}\label{fact0} 
$r$-genuine scaling dimensions belong to the set
\begin{align} \label{Set2}
    \Delta_{r-\text{genuine}} &\in  
    \left\{ \frac{d}{d- a} \ \bigg|  \ 0 < a \leq d-1 \ , \ \varphi(d) = 2r, \ \gcd(d,a) = 1 \right\}.
\end{align}
\end{property}
\begin{property}\label{fact10}
If a rank-$r$ $\mathcal{N}{=}2$ SCFT has a CB operator with an $\ell$-genuine scaling dimension $\Delta_i$, the associated $u_i$ coordinate axis $\mathcal{I}_i$ is not contained in an SK stratum of dimension less than $\ell$.
\end{property}

We will now show how Properties \ref{fact0} and \ref{fact10} follow from the $\mathbb{C}^\times$ symmetry action on the CB and the SK stratification.
Their derivations are closely related, and we will see that Property \ref{fact10} is almost a corollary of Property \ref{fact0}.

First, recall that the $\mathbb{C}^\times$ action on the CB is given by \eqref{C*action} and on the abelian fibers of the integrable system by \eqref{cE act fiber}.
If $\exp(z\mathcal{E})$ is in the stabilizer of $u \in \mathcal{C}$, i.e., if $\exp(z\mathcal{E}) \cdot u = u$, then by \eqref{cE act fiber} it defines an automorphism of the fiber, $\exp(z\mathcal{E}) \cdot X_u \cong X_u$.

Now consider the $u_i$ coordinate axis, $\mathcal{I}_i$, and consider a point on it, $u \in \mathcal{I}_i$.
Since $\mathcal{I}_i$ is a $\mathbb{C}^\times$ orbit, there is some value of $z$ such that $\exp(z\mathcal{E})$ fixes $u$.
Indeed, from its action \eqref{C*action}, it follows that
\begin{align}
    \xi^i \doteq \exp \left( \frac{2\pi i}{\Delta_i} \mathcal{E} \right)
\end{align}
generates the discrete subgroup of the $\mathrm{U}(1)_R$ symmetry which fixes $u$.
(Indeed, it fixes all of $\mathcal{I}_i$ pointwise.)
Furthermore it generates an automorphism of the fiber $X_u$.
From the action \eqref{C*action} of $\mathcal{E}$ on the $\mathcal{C}$ (the CB), it follows that $\xi^i$ acts on the CB cotangent space by
\begin{align}\label{act1}
    \diff u_{j} &\mapsto \xi^{i}(\diff u_{j}) = \exp{\left( \frac{2 \pi i}{\Delta_{i}}  \Delta_{j} \right) } \diff u_{j} .
\end{align}
Also, even though the fiber $X_u$ might be singular (if $\mathcal{I}_i$ belongs to an SK stratum), an appropriate notion of $\omega^{-1}$ still exists, as explained in the last section.
Furthermore, the \eqref{cE act fiber} action implies that $\omega^{-1}$ transforms with weight $\Delta_{\omega^{-1}} = -1$, so $\xi^i$ acts on it as
\begin{align}\label{act2}
    \omega^{-1} &\mapsto \xi^{i}(\omega^{-1}) = \exp{\left( -\frac{2 \pi i}{\Delta_{i}} \right) } \omega^{-1} .
\end{align}
By composing the above actions, we obtain the corresponding action on the tangent space of the (appropriate component of the perhaps singular) fiber, $T_\mu X_u$,
\begin{align}\label{eig1} 
    \omega^{-1} \circ \diff u_{j} &\mapsto \xi^{i}(\omega^{-1}) \circ \xi^{i}(\diff u_{j})
    =  \exp{\left( \frac{2 \pi i (\Delta_{j}-1)}{\Delta_{i}} \right) } (\omega^{-1} \circ \diff u_{j}),
\end{align}
for $j=1,\ldots,r$. 
Denoting this representation by $\widehat\rho_a(\xi^{i}) \in \mathrm{GL}(r, \mathbb{C})$, it has
\begin{align}\label{Rep1}
  \text{eigenvalues of } \widehat\rho_a(\xi^i) =
  \Bigl\{ \exp\bigl(2\pi i(\Delta_j{-}1)/\Delta_i \bigr), \ \ j = 1, \ldots, r \Bigr\}.
\end{align}

Now suppose the $u_i$ coordinate axis $\mathcal{I}_i$ belongs to a stratum of singular fibers of some dimension $0<\ell\le r$.
(The case $\ell = r$ means $\mathcal{I}_i$ does not belong to any stratum of singularities, i.e., is regular.)
By last section's discussion of the SK stratification, the singular fibers of an $\ell$-dimensional stratum have an $\ell$ complex dimensional abelian variety factor, $A_u$, and $\omega^{-1}$ maps cotangent directions to the stratum to tangent directions of $A_u$.
Since, by definition, $\diff u_i$ is cotangent to the stratum containing $\mathcal{I}_i$, we see that $\xi^i$ acts as an automorphism of $A_u$ with eigenvalue $\exp(-2\pi i/\Delta_i)$.
Note that $\ell-1$ other values taken from the set \eqref{Rep1} will also be eigenvalues of this automorphism, since the tangent space to $A_u$ is $\ell$-complex-dimensional.
In other words, the above $\widehat\rho_a(\xi^i)$ representation restricts to a representation
\begin{align}\label{anal rep}
    \rho_a \doteq \widehat\rho_a \Bigl\vert_{A_u}: \langle \xi_i\rangle \to \mathrm{GL}(\ell,\mathbb{C}),
\end{align}
of the cyclic subgroup of $\mathrm{Aut}(A_u)$ generated by $\xi_i$ acting on its tangent space $T_0 A_u$.

This fact places very strong constraints on the possible value of the scaling dimension $\Delta_i$ since the eigenvalues of automorphisms of abelian varieties can only belong to a small set of roots of unity.
We will review this argument \cite{Argyres:2018urp, Caorsi:2018zsq, Argyres:2018zay} now.

Note that any automorphism $\xi^i: A_u \to A_u$ of an abelian variety $A_u=\mathbb{C}^\ell/\Lambda$ can be uniquely lifted to a $\mathbb{C}$-linear map $\rho_a(\xi^i) \in \mathrm{GL}(\ell,\mathbb{C})$ on the covering space (which is the tangent space) with the property $\rho_a(\xi^i)(\Lambda) = \Lambda$. 
This gives the representation \eqref{anal rep} of the automorphism group.
On the other hand, we could equally consider the restriction $\rho_r(\xi^i) = \rho_a(\xi^i)\bigl\vert_\Lambda$ of $\rho_a$ to the lattice $\Lambda$ to get an integral representation
\begin{align}\label{eq:equiv}
    \rho_r: \mathrm{Aut}(A_u) \to \mathrm{Sp}_J(2\ell,\mathbb{Z}).
\end{align}
The image has to be in $\mathrm{Sp}_J(2\ell,\mathbb{Z})$ since an automorphism preserves not only the lattice but also its polarization $J$.
We call $\rho_a$ and $\rho_r$ the \emph{analytic representation} and \emph{rational representation} of $\mathrm{Aut}(A_u)$, respectively. 
These two representations are related by
\begin{align}\label{eq:equiv_reps}
    \rho_r\otimes_\mathbb{Q} \mathbb{C} \cong \rho_a \oplus \bar{\rho_a},
\end{align}
since if $A$, $R$ are matrices representing $\rho_{a}(\xi^i)$, $\rho_{r}(\xi^i)$, respectively, then they are related by the period matrix \eqref{per mat} by $A \Pi = \Pi R$. 
As $R$ is integer valued, complex conjugation gives $\bar{A}\, \bar{\Pi} = \bar{\Pi} R$, giving $\begin{psmallmatrix} A & 0\\ 0 & \bar{A} \end{psmallmatrix} \begin{psmallmatrix} \Pi \\ \bar{\Pi} \end{psmallmatrix} = \begin{psmallmatrix} \Pi \\ \bar{\Pi} \end{psmallmatrix} R$, which,  by \eqref{detP}, is an isomorphism \eqref{eq:equiv_reps} between the two representations.

Returning to our setup, \eqref{eq:equiv_reps} tells us that, when accompanied by the conjugate representation, the analytic representation \eqref{anal rep} is actually equivalent to an integral representation  $\rho_r(\xi^i) \in \mathrm{Sp}_J(2\ell,\mathbb{Z})$.
So, in particular, $\exp(-2\pi i/\Delta_i)$ must occur as the eigenvalue of an $\mathrm{Sp}_J(2\ell,\mathbb{Z})$ matrix all of whose eigenvalues are taken from the set \eqref{Rep1}.
This is very constraining due to the fact that the characteristic polynomial of an $\mathrm{Sp}_J(2\ell,\mathbb{Z})$ matrix with unit norm eigenvalues can be written in the form
\begin{align}\label{char poly}
    \mathrm{char}(\rho_r(\xi^i))(z) &= \prod_{k\ge1} \Phi_k(z)^{n_k}, & \sum_{k\ge1} \varphi(k) \, n_k &= 2\ell,
\end{align}
where $n_k$ are some non-negative integers, $\varphi$ is the Euler totient function, and $\Phi_\ell$ is the $k^{\text{th}}$ cyclotomic polynomial
\begin{align}
    \Phi_k(z) &\doteq \!\!\! \prod_{\substack{1\leq m \leq k \\ \gcd(k,m)=1}} \!\!\! (z - e^{2\pi i m/k}).
\end{align}
This follows because the characteristic polynomial of an integral matrix has integer coefficients, and the cyclotomic polynomials are the unique irreducible polynomials over the integers whose roots are roots of unity.
The second condition in \eqref{char poly} comes from the fact that $\mathrm{deg}(\Phi_k) = \varphi(k)$, and that $\mathrm{deg}(\mathrm{char}(\rho_r(\xi^i))) = 2\ell$.
It implies that the only cyclotomic polynomials that can appear in $\mathrm{char}(\rho_r(\xi^i))$ are those with
\begin{align}\label{tot bound}
    \varphi(k) \le 2\ell,
\end{align}
and so the eigenvalues of $\rho_r(\xi^i)$ are in the set
\begin{align}\label{Rep2}
  \text{eigenvalues of } \rho_r(\xi^i) \in
  \Bigl\{ \exp\bigl(2\pi i m/k \bigr), \ 1\le m\le k, \ \gcd(k,m) = 1, \ \varphi(k) \le 2\ell\ \Bigr\}.
\end{align}
Since $\exp(-2\pi i/\Delta_i)$ is in this set, and since, by unitarity, $\Delta_i >1$, we learn that $\Delta_i$ must belong to the set \eqref{Set} (with $r=\ell$).
In particular, the possible new scaling dimensions at rank $\ell$ are those that saturate totient bound \eqref{tot bound}, giving Property \ref{fact0}.
The result recorded in \eqref{Set2} is the subset of \eqref{Set} with $\varphi(n)=2r$, where we have reparameterized it in terms of $d \doteq n$ and $a \doteq n-m$ for later convenience.

Now suppose $\Delta_i$ is an $\ell$-genuine scaling dimension, so it is an element of the set \eqref{Set2} with $r=\ell$.
Then, if the $\mathcal{I}_i$ coordinate axis is contained in an SK stratum of dimension $k < \ell$, then the SK stratification implies $\exp(-2\pi i/\Delta_i)$ is an eigenvalue of an automorphism of a dimension-$k$ abelian variety fiber, while the above argument implies that such a $\Delta_i$ in the set \eqref{Set} with $r=k$, which does not include \eqref{Set2} with $r=\ell$.
This contradiction shows that $\mathcal{I}_i$ cannot be contained in any SK stratum of dimension $k < \ell$, thus showing Property \ref{fact10}.

Immediate consequences of Property \ref{fact10} are
\begin{property}\label{prop3} 
If a rank-$r$ $\mathcal{N}{=}2$ SCFT has a CB operator with an $r$-genuine scaling dimension $\Delta_i$, the associated $u_i$ coordinate axis $\mathcal{I}_i$ is necessarily non-singular,
\end{property}
\noindent and, with a bit more work, its generalization,
\begin{property}\label{prop4}
If an $\ell$-dimensional $\mathbb{C}^\times$-stratum, $\mathcal{I}_{i_1 \ldots i_\ell}$, is contained in an $\ell$-dimensional SK stratum, then its associated scaling dimensions are all allowed at rank $\ell$.
\end{property} 
\noindent This latter follows from the fact that if $\mathcal{I}_{i_1 \ldots i_\ell}$ is contained in an $\ell$-dimensional SK stratum, then a subset of $\ell$ of the eigenvalues in \eqref{Rep1} will be eigenvalues of an $\mathrm{Sp}_J(2\ell,\mathbb{Z})$ matrix.

These properties imply that subsets of $r$-tuples of scaling dimensions must satisfy the consistency conditions.
Define a \emph{genuine $\ell$-tuple} to be an $\ell$-tuple of $k$-genuine scaling dimensions with $k\le \ell$ and with a least one $\ell$-genuine scaling dimension saturating this inequality.
Then Properties \ref{fact0}--\ref{prop4} imply
\begin{property}\label{prop5}
    Let $\mathcal{D}=\{\Delta_1,\ldots,\Delta_r\}$ be an $r$-tuple of scaling dimensions of a rank-$r$ CB with $\Delta_i=d_i/(d_i-a_i)$ genuine in rank-$\ell_i$. If the abelian variety factor of the resolved fiber over $\mathcal{I}_i$ is of dimension $(\ell_i+q)$ with $0\leq q\leq r-\ell_i$, then $(\ell_i+q)$-many elements of the set
    \begin{gather}
        \{\exp(2\pi i (\Delta_j-1)/\Delta_i):\, j=1,\ldots,r\},
    \end{gather} 
    and their conjugates form the roots of a polynomial
    \begin{gather}
        P_i(z) = \Phi_{d_i}(z) \prod_{k\geq 1} \Phi_k(z)^{n_k},\quad\quad \sum_{k\geq 1} \varphi(k) n_k =2q.
    \end{gather}
\end{property}

\paragraph{Aside.} 
\label{non-totient}

Note that for certain ranks $r$, there are no $r$-genuine scaling dimensions, i.e., the set \eqref{Set2} may be empty.
This is due to the existence of {\it non-totient numbers}; that is, numbers $q$ for which no solution to $\varphi(p)=q$ exists. 
Of course, all odd numbers greater than $1$ are non-totient, but there are, in fact, infinitely many even non-totient numbers too. 
The first few are ({\tt \href{https://oeis.org/A005277}{A005277}} in OEIS):
\begin{gather}
    14,\ 26,\ 34,\ 38,\ 50,\ 62,\ 68,\ 74,\ 76,\ 86,\ 90,\ldots.
\end{gather}
Comparing with \eqref{Set2}, we see that SCFTs of rank-$7$ have no genuinely rank-$7$ scaling dimensions, for example. 
In these cases, a non-singular $\mathbb{C}^\times$ stratum in the CB need not exist.

\subsection{Constraints on $r$-genuine tuples}
\label{sec:conds}

We now use the properties derived in the previous subsection to constrain the sets of allowed $r$-tuples of CB scaling dimensions.
In this subsection we will focus on genuine $r$-tuples at rank $r$.
Recall that these are tuples of scaling dimensions of a rank-$r$ CB with at least one genuine rank-$r$ scaling dimension. 
Then, in the next subsection we will discuss the constraints coming from non-genuine scaling dimensions.

Since we have assumed that the given $r$-tuple is genuine, there is always (at least) one scaling dimension that is genuine; call it $\Delta_i$, and take $u\in \mathcal{I}_i$.
By property \ref{prop3}, the $u_{i}$-coordinate axis, $\mathcal{I}_{i}$, is necessarily non-singular. 
Since $\mathcal{I}_{i}$ is non-singular, its abelian fiber has dimension $r$, so $\widehat\rho_a(\xi^i)$ is the analytic representation of an automorphism of a dimension-$r$ abelian variety.
Thus, all its eigenvalues, \eqref{Rep1}, and by \eqref{eq:equiv_reps} their conjugates as well, are the eigenvalues of an $\mathrm{Sp}_J(2r,\mathbb{Z})$ matrix.
Thus the roots of its characteristic polynomial are exactly
\begin{align}
    \lambda_j^{\pm} &= \exp\left(\pm 2\pi i\frac{\Delta_j-1}{\Delta_i}\right), & j &= 1,\ldots,r .
\end{align}

Now recall that $\Delta_i$ is a genuinely rank-$r$ scaling dimension. 
Thus we can write, as in \eqref{Set2}, $\Delta_i = d_i/(d_i-a_i)$ for some coprime integers $a_i$ and $d_i$ satisfying $a_i\leq d_i-1$ and $\varphi(d_i)=2r$. 
As such,
\begin{gather}
    \lambda_i^{+} = \exp\left(-2\pi i \frac{d_i-a_i}{d_i}\right),
\end{gather}
is a root of $\mathrm{char}(\rho_r(\xi^i))$. 
Since this is a root of the cyclotomic polynomial $\Phi_{d_i}$, and since this polynomial has $2r$ distinct roots (the primitive $d_i$th roots of unity), we must have that the characteristic polynomial \eqref{char poly} is in fact $\mathrm{char}(\rho_r(\xi^i)) = \Phi_{d_i}$.
By comparing with the general form of a cyclotomic polynomial, we get that the other roots give rise to the relations
\begin{gather}\label{eq:delts}
    \frac{\Delta_j-1}{\Delta_i} = \frac{a_j}{d_i} \ \ \Longrightarrow\ \ \Delta_j =1+\frac{a_j}{d_i-a_i}, \qquad \forall j \neq i,
\end{gather}
for some other integers $a_j$ coprime to $d_i$.

Notice that \eqref{eq:delts} must be satisfied for all genuine scaling dimensions $\Delta_i$ against all other scaling dimensions $\Delta_j$ in the candidate $r$-tuple. 
The set of relations \eqref{eq:delts} may be satisfied for one genuine scaling dimension $\Delta_i$, but the corresponding set of relations may fail to be satisfied relative to a different genuine scaling dimensions in the $r$-tuple (should one exist).
We call this feature the \emph{non-reflexivity} of the genuine scaling dimension condition --- see, e.g., example \ref{ex:18/17} below.

The fact that the roots of $\mathrm{char}(\rho_r(\xi^i)) = \Phi_{d_i}$ are all distinct has several non-trivial consequences, which we now outline.
\begin{enumerate}
    \item Since each $\lambda^\pm_j$ must correspond to a different root of unity, we get the modular constraint
    \begin{gather}\label{eq:unity}
        a_j-a_k \neq 0 \modulo{d_i}, \quad\forall j\neq k.
    \end{gather}
    \item The eigenvalues of an $\mathrm{Sp}_J(2r,\mathbb{Z})$ matrix come in reciprocal pairs $\{\lambda_j^+,\lambda_j^-\}$. These reciprocal pairs signal the existence of a single CB scaling dimension, so we must take care to not count these as separate putative CB dimensions. This gives the constraint
    \begin{gather}\label{eq:recip}
        a_j+a_k \neq 0 \modulo{d_i}, \quad \forall j\neq k.
    \end{gather}
    \item An immediate consequence of the previous points is that the presence of an $r$-genuine CB dimension forbids repeated dimensions. Repeated dimensions can, however, occur in non-genuinely rank-$r$ tuples.
    \item 
    Another interesting corollary of \eqref{eq:recip} is that we cannot have a CB scaling dimension $\Delta_j=2$ in genuine $r$-tuples. Indeed, this would correspond to the eigenvalue
    \begin{gather}
        \lambda_i^- = \exp\left(2\pi i \frac{1}{\,\Delta_i}\right),
    \end{gather}
    but this is the reciprocal of the eigenvalue corresponding to $\Delta_i$, which is genuinely rank-$r$, leading to a contradiction.
    \item 
    Also, if we take $\Delta_i$ to be an integer genuinely rank-$r$ scaling dimension, equation \eqref{eq:delts} tells us that all other dimensions are also integer. Furthermore, if $\Delta_i$ is even, then so are the other compatible dimensions.
\end{enumerate}
To summarize, equations \eqref{eq:delts}, \eqref{eq:unity}, and \eqref{eq:recip} are our main results constraining the possible sets of genuinely rank-$r$ tuples.

\medskip
 
To illustrate these constraints, let us now consider some low rank examples.

\begin{example}\label{ex:8/7}
    Consider the genuinely rank-$2$ scaling dimension $\Delta_1=\tfrac{8}{7}$. In order for another CB scaling dimension $\Delta_2$ to be consistent, we must have
    \begin{gather}
        \Delta_2 = 1 + \frac{a_2}{7},\ \ \gcd(a_2,8)=1.
    \end{gather}
    From this we see that $a_2\, (\mathrm{mod}\ 8)\in\{1,3,5,7\}$. However, $\Delta_1$ corresponds to $a_1=1$, meaning that $a_2 = 7\, (\mathrm{mod}\ 8)$ is ruled out due to equation \eqref{eq:recip} and $a_2  = 1\, (\mathrm{mod}\ 8)$ is also ruled out due to equation \eqref{eq:unity}. 
    For $a_2=3 \modulo{8}$, only $a_2=3$ and $35$ give a $\Delta_2$ in the set of allowed rank-2 dimensions \eqref{eq:rank2}.  Likewise, for $a_2=5 \modulo{8}$, only $a_2=5$ and $21$ are allowed.
    This leaves us with the only possible pairs $\{\tfrac{8}{7},\tfrac{10}{7}\}$, $\{\tfrac{8}{7},\tfrac{12}{7}\}$, $\{\tfrac{8}{7},4\}$ and $\{\tfrac{8}{7},6\}$. 
    The first two pairs each involve a second genuinely rank-2 scaling dimension, so they must be checked against the constraints \eqref{eq:delts}--\eqref{eq:recip}, but now with $\Delta_i = \frac{10}{7}$ and $\frac{12}{7}$, respectively.  In both these cases it is easy to see that the constraints are satisfied.
    Indeed, the $\{\tfrac{8}{7},\tfrac{10}{7}\}$ pair are the scaling dimensions of the $(A_1,A_4)$ Argyres-Douglas theory, which, to the best of our knowledge, is the only known absolute rank-$2$ SCFT with two genuinely rank-$2$ scaling dimensions.\footnote{It is interesting to remark that the $(A_1,D_7)$ theory \cite{Eguchi:1996ds} has CB operators with scaling dimensions $\{8/7,10/7,12/7\}$, computed using the methods of \cite{Shapere:1999xr}, thus giving an irreducible rank-3 SCFT with only 2-genuine scaling dimensions.}
\end{example} 

\begin{example}\label{ex:18/17}
    Consider the genuinely rank-$3$ scaling dimension $\Delta_1=\tfrac{18}{17}$, so $d_1=18$ and $a_1=1$. 
    The consistent triplets of CB scaling dimensions $\{\Delta_1,\Delta_2,\Delta_3\}$ must satisfy
    \begin{gather}
        \Delta_j = 1 + \frac{a_j}{17},\ \ \gcd(a_j,18)=1, \ \ j\in\{2,3\}.
    \end{gather}
    As such, we must have $a_j\, (\mathrm{mod}\ 18) \in \{1,5,7,11,13,17\}$. 
    The only solutions to these equations consistent with the set of rank-$3$ scaling dimensions are $\{85,119,187,221,289\}$ which correspond to scaling dimensions $\{6,8,12,14,18\}$ respectively. 
    Na\"ively, one could say that there are, therefore, ${5 \choose 2}$-many possible sets of CB dimensions including $\tfrac{18}{17}$. However, checking the modular constraints shows that this is not the case. 
    Indeed, $119+187 = 0 \modulo{18}$ and $85+221 = 0 \modulo{18}$, both violating equation \eqref{eq:recip}, while $289-1=0 \modulo{18}$ violates equation \eqref{eq:unity}. We thus conclude that the only triplets consistent with $\Delta_1=\tfrac{18}{17}$ are given by $\{\tfrac{18}{17}, 6, 12\}$, $\{\tfrac{18}{17},6,8\}$, $\{\tfrac{18}{17},8,14\}$ and $\{\tfrac{18}{17}, 12,14\}$. However, as $14$ is also genuinely rank-$3$, we must check that the latter two triplets are consistent with our construction, but now with $\Delta_1=14$. Doing so shows that $\tfrac{18}{17}$ cannot be present when $14$ is, therefore leaving $\{\tfrac{18}{17}, 6, 12\}$ and $\{\tfrac{18}{17},6,8\}$ as the only truly valid triplets.
\end{example}

\subsection{Constraints on non-genuine \texorpdfstring{$r$}{r}-tuples.}
\label{sec non-gen}

Key to our discussion in the genuine case was the fact that the $u_i$-plane $\mathcal{I}_i$ was non-singular. 
This can no longer be guaranteed in the non-genuine case, as any $(r-1)$-subtuple is allowed at rank-$(r-1)$. 
Nevertheless, the scaling dimensions are still constrained by Properties \ref{fact0}--\ref{prop5}, derived above. 

Recall that if $u_i$ has an $\ell$-genuine scaling dimension, Property \ref{fact10} tells us that its corresponding coordinate axis is not contained in an SK stratum of dimension less than $\ell$. In turn, this means that the resolved fiber over this axis contains an abelian variety $A$ of dimension at least $\ell$. We stress that it is the dimension of this variety, and not just the value of $\ell$, that is important in constraining the tuple of scaling dimensions. Indeed, this is showcased by Property \ref{prop5}, which states that exactly $\dim A\geq \ell$ many scaling dimensions are constrained. The following example highlights this fact.

\begin{example}\label{ex:classS}
    Consider the six dimensional type $A_{9}$ $\mathcal{N}=(2,0)$ theory compactified on a sphere with one maximal puncture, labeled by the partition $[1^{10}]$, and two non-maximal punctures labeled by $[5^2]$ and $[4^2,2]$.\footnote{This theory was recently discussed in \cite{Giacomelli:2024dbd}. We thank the authors for pointing out this example to us.} By using the rules outlined in \cite{Chacaltana:2010ks}, one obtains a Hitchin field with polar orders
    \begin{equation}
        \begin{aligned}
            p_1=(0,1,2,3,4,5,6,7,8,9),\\
            p_2=(0,1,1,2,2,3,3,4,4,5),\\
            p_3=(0,1,2,2,3,4,4,5,5,6),
        \end{aligned}
    \end{equation}
    from which we find Coulomb branch operators $\{u_1,u_2,u_3\}$ with scaling dimensions $\{6,8,10\}$ respectively. Furthermore, the Seiberg-Witten curve of this theory is given by 
    \begin{gather}\label{eq:swcurve}
        \lambda^{10}-u_1\phi_6(y) \lambda^4-u_2 \phi_8(y)\lambda^2-u_3\phi_{10}(y)=0,
    \end{gather}
    where $\lambda = x\diff y$ and $\phi_k(y)$ are $k$-differentials. Regarding this as a polynomial equation for $x$, we can now analyze how the curve behaves at the coordinate axes $\mathcal{I}_i$.
    
    \medskip
    \noindent {\bf Axis $\mathcal{I}_3$.} When $u_1=u_2=0$ and $u_3\neq 0$, the curve is generically non-singular. As such, the corresponding abelian fiber is three dimensional and possesses complex multiplication by $\exp(-2\pi i/10)$. The set \eqref{Rep1} is given by
    \begin{gather}\label{eq:roots}
        \{ -1, \exp(-3\pi i/5), \exp(-\pi i /5)\},
    \end{gather}
    which, together with their conjugates, forms the roots of the polynomial
    \begin{gather}
        P_3(z) = \Phi_{10}(z)\Phi_2(z)^2.
    \end{gather}
    Note that despite $\Delta=10$ being 2-genuine, we still get a constraint on all the dimensions of the triple since the abelian variety is smooth over $\mathcal{I}_3$. This will not be in the case for $\Delta=8$, as we will show now.
    
    \medskip
    \noindent {\bf Axis $\mathcal{I}_2$.} By taking $u_3=0$, we see that \eqref{eq:swcurve} develops two coincident roots, so this locus belongs to the discriminant locus and supports a rank-2 special geometry. There is an easy candidate for this rank-2 theory; consider the class $\mathcal{S}$ theory of type $A_7$ on a sphere with three punctures with polar orders
    \begin{equation}
        \begin{aligned}
            p_1=(0,1,2,3,4,5,6,7),\\
            p_2=(0,1,1,2,2,3,3,4),\\
            p_3=(0,1,2,2,3,4,4,5),
        \end{aligned}
    \end{equation}
   which gives a theory with Coulomb branch dimensions $\{6,8\}$. The reduced polynomial
   \begin{gather}
       \lambda^8-u_1 \phi_6(y)\lambda^2-u_2\phi_8(y)=0,
   \end{gather}
    has non-zero discriminant for generic $u_1$ and $u_2$. As such, the abelian variety fibered over $\mathcal{I}_2$ has dimension 2 and possesses complex multiplication by $\exp(-2\pi i/8)$. The set \eqref{Rep1} is given by
    \begin{gather}
        \{\exp(-6\pi i/8),\, \exp(-2\pi i/8),\, \exp(2\pi i/8)\}.
    \end{gather}
    The first two entries and their conjugates form the roots of $\Phi_8(z)$, in agreement with proposition \ref{prop5}.
    
    \medskip
    \noindent {\bf Axis $\mathcal{I}_1$.} Finally, when $u_2=u_3=0$ and $u_1\neq 0$, the curve \eqref{eq:swcurve} has 4 coincident roots, signaling that a rank-1 special geometry is supported along $\mathcal{I}_1$. More precisely, the theory supported along this stratum is the class $\mathcal{S}$ theory of type $A_5$ with three punctures of polar orders
    \begin{equation}
        \begin{aligned}
            p_1=(0,1,2,3,4,5),\\
            p_2=(0,1,1,2,2,3),\\
            p_3=(0,1,2,2,3,4),
        \end{aligned}
    \end{equation}
    which is a rank-1 SCFT with $\Delta=6$. The characteristic polynomial in this case is $\Phi_6(z)$, which is of degree two.
\end{example}

We saw in that in the genuine case, no scaling dimensions could repeat due to the fact that all the roots of $\Phi_{d_i}(z)$ are distinct. However, in the non-genuine case this can occur. Nonetheless, the form of the characteristic polynomial \eqref{char poly} and Property \ref{prop5} give the following constraint on repeating dimensions.

\begin{property}\label{prop6}
    Let $\{\Delta_1,\ldots,\Delta_r\}$ be an $r$-tuple of scaling dimensions. Suppose that $\Delta_{i}$ repeats $s$-many times in the $r$-tuple. Then $\Delta_i$ is an allowed dimension in rank $\lfloor r/s\rfloor$.\footnote{Note the arguments of \cite{Cecotti:2021ouq} generalize to each stratum. That is, one can define a ``characteristic dimension'' associated to each stratum that will take values in the set of scaling dimensions of rank $\lfloor r/s\rfloor$.}
\end{property}
\noindent In particular, if $r=s$ we see that the only possibility is that $\Delta_i$ is rank-1.

In the next section we outline a systematic procedure to generate all allowed $r$-tuples, whether genuine or not.

\section{Allowed \texorpdfstring{$r$}{r}-tuples of CB scaling dimensions}\label{sec4}

We present an algorithm for constructing the allowed $r$-tuples of CB operator scaling dimensions at a given rank $r$. 
Our algorithm considers the genuine rank-$r$ tuples, and non-genuine rank-$r$ tuples separately. 

\paragraph{Genuine $r$-tuples.}

To calculate the allowed genuinely rank-$r$ tuples, we generalize the procedures outlined in examples \ref{ex:8/7} and \ref{ex:18/17}.
\begin{enumerate}
    \item Select a genuine rank-$r$ scaling dimension.
    \item Using \eqref{eq:delts} and the relations between the $\{a_j\}$ calculate which of the elements of \eqref{Set} can occupy a tuple with the chosen genuine dimension.
    \item Generate all possible $r$-tuples of the genuine dimension with the dimensions it can occupy a tuple with.
    \item Note that if a tuple contains more than 1 genuinely rank-$r$ dimension, then we must check our consistency conditions with all of them to ensure the tuple is consistent. 
    \item For each non-genuine scaling dimension in the tuple, check that Property \ref{prop5} is satisfied. If this is the case, then the resulting tuple is a valid genuine tuple. 
\end{enumerate}
Repeating this process for all of the genuine rank-$r$ scaling dimensions completely determines the set of genuinely rank-$r$ tuples. 
In appendix \ref{app:tables} we present the results of this algorithm for rank-$2$, $3$ and $4$. 

\paragraph{Non-genuine $r$-tuples.}

In the previous case, the genuine scaling dimensions constrained every element of the tuple, thus allowing us to easily construct all allowable genuine tuples. Without a genuine scaling dimension, the best that we can do is implement Property \ref{prop5} on every stratum. By assuming that the dimension of the resolved abelian variety factor above that stratum is of minimum dimension, we can then generate all possible non-genuine tuples.

\begin{enumerate}
    \item Select a rank-$\ell_1$ scaling dimension $\Delta_1$ with $1\leq \ell_1 < r$ and form an $\ell_1$-tuple $\mathcal{D}$ that satisfies Property \ref{prop5} including $\Delta_1$. 
    \item Select a rank-$\ell_2$ scaling dimension $\Delta_2$ and form an $\ell_2$-tuple $\mathcal{D}'$ satisfying Property \ref{prop5}. Let $\mathcal{D}''$ be the union of $\mathcal{D}$ from the previous step and $\mathcal{D}'$ allowing for repetition of elements.
    \item If any element of $\mathcal{D}''$ that is repeated $t$-many times was present in both $\mathcal{D}$ and $\mathcal{D}'$, choose whether or not to erase up to $(t-1)$-many copies of that element. Redefine $\mathcal{D}$ to be $\mathcal{D}''$ after this reduction.
    \item Repeat steps 2--3 until an $r$-tuple is formed and check Property \ref{prop5} is satisfied for any scaling dimensions not selected at the start of steps 1 and 2. If Property \ref{prop6} is also satisfied, the resulting tuple is a valid non-genuine tuple at rank-$r$.
\end{enumerate}

\noindent To illustrate the the algorithm, let us describe the possibilities at low rank:
\begin{itemize}
 \item For rank-$2$ theories, the only option of a non-genuine pair is to take two rank-$1$ scaling dimensions, to form a rank-$2$ pair. These pairs are realized by the tensor product of two rank-$1$ theories.
 \item For rank-$3$, we again have the tuples realized by product theories-- triples of rank-1 dimensions or the union of a genuine rank-$2$ pair with a rank-$1$ dimension. We also can consider two pairs $\{\Delta_1,\Delta_2\}$ and $\{\Delta_2,\Delta_3\}$ and apply our algorithm to get the triple $\{\Delta_1,\Delta_2, \Delta_3\}$. Note that each of the individual pairs need \emph{not} be the scaling dimensions of a rank-$2$ theory, as long as Property \ref{prop5} is satisfied. This occurs in Example \ref{ex:classS}, where neither $\{6,10\}$ nor $\{8,10\}$ are valid rank-$2$ pairs, but equation \eqref{eq:roots} shows that Property \ref{prop5} is still satisfied. The pairs of rank-$2$ dimensions of this type are 
\begin{gather*}
         \left\{\tfrac{12}{11},8\right\},
\left\{\tfrac{10}{9},8\right\},
\left\{\tfrac{8}{7},12\right\},
\left\{\tfrac{5}{4},8\right\},
\left\{\tfrac{10}{7},8\right\},\\
\left\{\tfrac{8}{5},\tfrac{12}{5}\right\},
\left\{\tfrac{8}{5},12\right\},
\left\{\tfrac{5}{3},8\right\},
\left\{\tfrac{12}{7},8\right\},
\left\{\tfrac{12}{5},8\right\},
\left\{\tfrac{5}{2},8\right\},\\
\left\{\tfrac{8}{3},\tfrac{10}{3}\right\},
\left\{\tfrac{8}{3},12\right\},
\left\{\tfrac{10}{3},8\right\},
\left\{5,8\right\},
\left\{8,10\right\}.
 \end{gather*}
One can easily see that these pairs do not pass our consistency conditions for one of the two genuine scaling dimensions, thus ruling them out as genuine pairs. Nonetheless, they can still arise on a stratum of a higher rank theory.
\end{itemize}

\noindent Of course, the above algorithm generates all allowed non-genuine $r$-tuples, but it is still possible there is a finer classification: we expect that the collection of actually realized scaling dimensions in 4d $\mathcal N=2$ SCFTs is a subset of the list generated by the algorithm above. 

\paragraph{Comparison with previous estimates.}

To round out this section, let us compare our new estimates for the number of rank-$r$ tuples with the na\"ive estimates using the results of \cite{Caorsi:2018zsq,Argyres:2018urp}.
As the rank grows large, the number of allowable rank-$r$ dimensions scales asymptotically as \cite{Caorsi:2018zsq}
\begin{gather}
    N_r = \frac{2\zeta(2)\zeta(3)}{\zeta(6)}r^2 + o(r^2).
\end{gather}
If we assume that any combination of scaling dimensions constitutes a valid rank-$r$ tuple, a standard stars and bars argument gives the na\"ive number of distinct $r$-tuples 
\begin{gather}
    T_r^\star = {N_r+r-1 \choose r}
    \sim \mathcal{O}\left( \frac{r^{2r}}{r!}\right).
\end{gather}
Included in this count is the approximate number of non-genuine tuples given by
\begin{gather}
    L^\star_r = {N_{r-1}+r-1 \choose r}.
\end{gather}
As such, the difference $T_r^\star-L_r^\star$ gives a rudimentary counting of the number of genuine $r$-tuples. By imposing our constraints on this putative set of scaling dimensions, we can see that this is a vast overestimate. 
For ranks less than or equal to $4$, we tabulate our results in table \ref{tab:numerics}.

\begin{table}
    \centering
    \begin{tabular}{cccccc} \hline\hline
        rank & $N_r$ & $T_r^\star$ & $L_r^{\star}$ & $T_r^{\mathrm{gen}}$ & $Q_r$ \\ \hline
        $2$ & $23$ &$276$& $28$ & $35$ & $1.41129\times 10^{-1}$\\
        $3$ & $47$ &$18424$ & $2300$ & $138$ & $8.55867\times 10^{-3}$\\
        $4$ & $87$ &$2555190$ & $230300$ & $216$ & $9.29076\times 10^{-5}$\\
        \hline\hline
    \end{tabular}
    \caption{A comparison of our estimates of the number of genuinely rank-$r$ tuples to the na\"ive counting. Here $T_r^{\mathrm{gen}}$ is the number of genuinely rank-$r$ tuples consistent with our constraints and $Q_r=T_r^{\mathrm{gen}}/(T_r^\star-L_r^\star)$ is the ratio of our estimates by the previous estimate on the number of genuine $r$-tuples.}
    \label{tab:numerics}
\end{table}

It would be useful and informative to obtain precise bounds on the growth of the number of allowed $r$-tuples, and on the statistics of their distribution.
See, for instance, \cite{Perlmutter:2024noo} for an example of an interesting application of this kind of information.

\begin{acknowledgments}
We thank Simone Giacomelli and Raffaele Savelli for helpful discussions regarding non-genuine tuples. PCA is supported in part by DOE grant DE-SC1019775. The work of MDZ has received funding from the European Research Council (ERC) under the European Union’s Horizon 2020 research and innovation program (grant agreement No. 851931), the VR project grant No. 2023-05590 and from the Simons Foundation (grant \#888984, Simons Collaboration on Global Categorical Symmetries). MDZ also acknowledges the VR Centre for Geometry and Physics (VR grant No. 2022-06593). The work of BS and MM is supported by STFC grant ST/T000759/1. RM is supported by a Knut and Alice Wallenberg Foundation postdoctoral scholarship in mathematics.
\end{acknowledgments}

\appendix

\section{Tables of genuine \texorpdfstring{$r$}{r}-tuples}
\label{app:tables}

In table \ref{tab:rank2} we present the valid genuinely rank-2 pairs. This corrects the findings of \cite{Kaidi:2022sng} where 16 additional pairs were reported. 
These additional 16 pairs had two genuinely rank-2 scaling dimensions that are ruled out by checking our compatibility conditions using \emph{both} of the scaling dimensions. 
For example, the pair $\left\{\tfrac{12}{5},8\right\}$ looks valid if we only check the compatibility conditions for $\tfrac{12}{5}$ but, as mentioned in section \ref{sec:conds}, the only dimensions valid with $8$ are even integers. 
As such, this is an invalid pair.

We present the genuinely rank-3 triplets in tables \ref{tab:rank3_part1} and \ref{tab:rank3_part2}, and the genuinely rank-4 quadruplets in tables \ref{tab:rank4_part1} -- \ref{tab:rank4_part3}. 
For tuples that contain more than one genuinely rank-$r$ scaling dimension, we have highlighted them in \blue{blue} if they have already appeared in the table for a smaller scaling dimension $\Delta_{\mathrm{new}}$.

\begin{table}
\begin{center}
\begin{tabular}{cc}
\hline\hline
$d$ &  $\{\Delta_1 , \Delta_2 \}$ \\  \hline 

\multirow{2}{*}{5} & $\left\{\tfrac{5}{4}, \tfrac{3}{2}\right\}$, $\left\{\tfrac{5}{4}, 3\right\}$, $\left\{\tfrac{5}{4}, 4\right\}$, $\left\{\frac{4}{3}, \tfrac{5}{3}\right\}$, $\left\{\tfrac{5}{3}, 3\right\}$, $\left\{\tfrac{5}{3}, 4\right\}$, $ \left\{\tfrac{3}{2}, \tfrac{5}{2}\right\}$, \\ & $\left\{\tfrac{5}{2}, 3\right\}$, $\left\{\tfrac{5}{2}, 4\right\}$, $\left\{3, 5\right\}$, $\{4, 5\}$ \\ \hline

\multirow{2}{*}{8} & $\left\{\tfrac{8}{7},\tfrac{10}{7} \right\}$, $\left\{\tfrac{8}{7},\tfrac{12}{7} \right\}$, $\left\{\tfrac{8}{7}, 4\right\}$, $\left\{\tfrac{8}{7}, 6\right\}$, $\left\{\tfrac{6}{5}, \tfrac{8}{5}\right\}$, $\left\{\tfrac{8}{5}, 4\right\}$, $\left\{\tfrac{8}{5}, 6\right\}$\\  & $\left\{\tfrac{4}{3}, \tfrac{8}{3}\right\}$, $\left\{\tfrac{8}{3}, 4\right\}$, $\left\{\tfrac{8}{3}, 6\right\}$, $\{4, 8\}$, $\{6, 8\}, \{8,12\}$ \\ \hline 

10 &  $\left\{\tfrac{10}{9}, \tfrac{4}{3}\right\}$, $\left\{\tfrac{10}{9}, 4\right\}$, $\left\{\tfrac{10}{7}, 4\right\}$, $\left\{\tfrac{4}{3}, \tfrac{10}{3}\right\}$, $\left\{\tfrac{10}{3}, 4\right\}$, $\{4, 10\}$ \\  \hline 

12 & $\left\{\tfrac{12}{11}, 6\right\}$, $\left\{\tfrac{12}{7}, 6\right\}$, $\left\{\tfrac{6}{5}, \tfrac{12}{5}\right\}$, $\left\{\tfrac{12}{5}, 6\right\}$, $\{6, 12\}$ \\  \hline\hline
\end{tabular}%
\caption{ The 35 genuine rank-$2$ scaling dimension pairs.}\label{tab:rank2}
\end{center}
\end{table}

\begin{sidewaystable}[t!]
\centering
\begin{tabular}{cc}\hline\hline
    $\Delta_{\mathrm{new}}$ & $\{\Delta_1,\Delta_2,\Delta_3\}$ \\ \hline 
            
    $\tfrac{18}{17}$& $\left\{\tfrac{18}{17},6,12\right\},\left\{\tfrac{18}{17},6,8\right\}$ \\ \hline
             
    $\tfrac{14}{13}$ & $\left\{\tfrac{14}{13},4,6\right\},
\left\{\tfrac{14}{13},4,10\right\},
\left\{\tfrac{14}{13},6,12\right\}$ \\ \hline
             
    {$\tfrac{9}{8}$} & $\left\{\tfrac{9}{8},\tfrac{5}{4},\tfrac{3}{2}\right\},
\left\{\tfrac{9}{8},\tfrac{3}{2},3\right\},
\left\{\tfrac{9}{8},3,5\right\},
\left\{\tfrac{9}{8},3,6\right\},
\left\{\tfrac{9}{8},6,8\right\},
\left\{\tfrac{9}{8},6,12\right\}$\\ \hline 
             
    \multirow{2}{*}{$\tfrac{7}{6}$} & $\left\{\tfrac{7}{6},\tfrac{4}{3},\tfrac{3}{2}\right\},
\left\{\tfrac{7}{6},\tfrac{4}{3},\tfrac{5}{3}\right\},
\left\{\tfrac{7}{6},\tfrac{4}{3},\tfrac{8}{3}\right\},
\left\{\tfrac{7}{6},\tfrac{4}{3},4\right\},
\left\{\tfrac{7}{6},\tfrac{3}{2},\tfrac{5}{2}\right\},
\left\{\tfrac{7}{6},\tfrac{3}{2},3\right\},
\left\{\tfrac{7}{6},\tfrac{3}{2},6\right\},
\left\{\tfrac{7}{6},\tfrac{5}{3},3\right\},
\left\{\tfrac{7}{6},\tfrac{5}{2},4\right\},$\\ &$
\left\{\tfrac{7}{6},\tfrac{8}{3},6\right\},
\left\{\tfrac{7}{6},3,4\right\},
\left\{\tfrac{7}{6},3,5\right\},
\left\{\tfrac{7}{6},4,6\right\},
\left\{\tfrac{7}{6},4,10\right\},
\left\{\tfrac{7}{6},6,12\right\},$\\ \hline
             
    $\tfrac{14}{11}$ & $\left\{\tfrac{14}{11},4,6\right\},
\left\{\tfrac{14}{11},4,10\right\},
\left\{\tfrac{14}{11},6,12\right\}$ \\ \hline
             
    {$\tfrac{9}{7}$} & $\left\{\tfrac{8}{7},\tfrac{9}{7},\tfrac{12}{7}\right\},
\left\{\tfrac{9}{7},\tfrac{12}{7},6\right\},
\left\{\tfrac{9}{7},3,5\right\},
\left\{\tfrac{9}{7},3,6\right\},
\left\{\tfrac{9}{7},6,8\right\},
\left\{\tfrac{9}{7},6,12\right\}$ \\ \hline
             
    $\tfrac{18}{13}$ & $\left\{\tfrac{18}{13},6,8\right\},\left\{\tfrac{18}{13},6,12\right\}$ \\ \hline
             
    \multirow{2}{*}{$\tfrac{7}{5}$} & $\left\{\tfrac{6}{5},\tfrac{7}{5},\tfrac{8}{5}\right\},
\left\{\tfrac{6}{5},\tfrac{7}{5},\tfrac{9}{5}\right\},
\left\{\tfrac{6}{5},\tfrac{7}{5},3\right\},
\left\{\tfrac{6}{5},\tfrac{7}{5},6\right\},
\left\{\tfrac{7}{5},\tfrac{8}{5},4\right\},
\left\{\tfrac{7}{5},3,4\right\}$\\ &$
\left\{\tfrac{7}{5},3,5\right\},
\left\{\tfrac{7}{5},4,6\right\},
\left\{\tfrac{7}{5},4,10\right\},
\left\{\tfrac{7}{5},6,12\right\}$\\ \hline
             
    {$\tfrac{14}{9}$} & $\left\{\tfrac{10}{9},\tfrac{4}{3},\tfrac{14}{9}\right\},
\left\{\tfrac{4}{3},\tfrac{14}{9},\tfrac{8}{3}\right\},
\left\{\tfrac{4}{3},\tfrac{14}{9},4\right\},
\left\{\tfrac{14}{9},\tfrac{8}{3},6\right\},
\left\{\tfrac{14}{9},4,6\right\},
\left\{\tfrac{14}{9},4,10\right\},
\left\{\tfrac{14}{9},6,12\right\}$ \\ \hline
             
    $\tfrac{18}{11}$ & $\left\{\tfrac{18}{11},6,8\right\},
\left\{\tfrac{18}{11},6,12\right\}$ \\ \hline \hline
\end{tabular}
\caption{The genuinely rank-$3$ sets of scaling dimensions with $\tfrac{18}{17}\leq \Delta_{\mathrm{new}}\leq \tfrac{18}{11}$.}
\label{tab:rank3_part1}
\end{sidewaystable}

\newpage
\begin{sidewaystable}[t!]
\centering
\begin{tabular}{cc}\hline\hline
    $\Delta_{\mathrm{new}}$&  $\{\Delta_1,\Delta_2,\Delta_3\}$ \\ \hline
         
    \multirow{2}{*}{$\tfrac{7}{4}$}& $\left\{\tfrac{5}{4},\tfrac{3}{2},\tfrac{7}{4}\right\},
\left\{\tfrac{5}{4},\tfrac{7}{4},4\right\},
\left\{\tfrac{3}{2},\tfrac{7}{4},\tfrac{5}{2}\right\},
\left\{\tfrac{3}{2},\tfrac{7}{4},3\right\},
\left\{\tfrac{3}{2},\tfrac{7}{4},6\right\},
\left\{\tfrac{7}{4},\tfrac{5}{2},4\right\},
\left\{\tfrac{7}{4},3,4\right\}$\\ &$
\left\{\tfrac{7}{4},3,5\right\},
\left\{\tfrac{7}{4},4,6\right\},
\left\{\tfrac{7}{4},4,10\right\},
\left\{\tfrac{7}{4},6,12\right\}$ \\ \hline
         
   $\tfrac{9}{5}$ & $\blue{\left\{\tfrac{6}{5},\tfrac{7}{5},\tfrac{9}{5}\right\}},
\left\{\tfrac{6}{5},\tfrac{9}{5},\tfrac{12}{5}\right\},
\left\{\tfrac{6}{5},\tfrac{9}{5},6\right\},
\left\{\tfrac{9}{5},3,5\right\},
\left\{\tfrac{9}{5},3,6\right\},
\left\{\tfrac{9}{5},6,8\right\},
\left\{\tfrac{9}{5},6,12\right\}$\\ \hline
         
    $\tfrac{9}{4}$ &  $\left\{\tfrac{5}{4},\tfrac{3}{2},\tfrac{9}{4}\right\},
\left\{\tfrac{3}{2},\tfrac{9}{4},3\right\},
\left\{\tfrac{9}{4},3,5\right\},
\left\{\tfrac{9}{4},3,6\right\},
\left\{\tfrac{9}{4},6,8\right\},
\left\{\tfrac{9}{4},6,12\right\}$ \\ \hline
         
    \multirow{2}{*}{$\tfrac{7}{3}$} & $\left\{\tfrac{4}{3},\tfrac{5}{3},\tfrac{7}{3}\right\},
\left\{\tfrac{4}{3},\tfrac{7}{3},\tfrac{8}{3}\right\},
\left\{\tfrac{4}{3},\tfrac{7}{3},4\right\},
\left\{\tfrac{5}{3},\tfrac{7}{3},3\right\},
\left\{\tfrac{5}{3},\tfrac{7}{3},6\right\},
\left\{\tfrac{7}{3},\tfrac{8}{3},6\right\}$\\ &$
\left\{\tfrac{7}{3},3,4\right\},
\left\{\tfrac{7}{3},3,5\right\},
\left\{\tfrac{7}{3},4,6\right\},
\left\{\tfrac{7}{3},4,10\right\},
\left\{\tfrac{7}{3},6,12\right\}$\\ \hline
         
    $\tfrac{18}{7}$ & $\left\{\tfrac{8}{7},\tfrac{12}{7},\tfrac{18}{7}\right\},
\left\{\tfrac{12}{7},\tfrac{18}{7},6\right\},
\left\{\tfrac{18}{7},6,8\right\},
\left\{\tfrac{18}{7},6,12\right\}$\\ \hline
         
    $\tfrac{14}{5}$ &  $\left\{\tfrac{6}{5},\tfrac{8}{5},\tfrac{14}{5}\right\},
\left\{\tfrac{6}{5},\tfrac{14}{5},6\right\},
\left\{\tfrac{8}{5},\tfrac{14}{5},4\right\},
\left\{\tfrac{14}{5},4,6\right\},
\left\{\tfrac{14}{5},4,10\right\},
\left\{\tfrac{14}{5},6,12\right\}$ \\ \hline
         
    \multirow{2}{*}{$\tfrac{7}{2}$} & $\left\{\tfrac{3}{2},\tfrac{5}{2},\tfrac{7}{2}\right\},
\left\{\tfrac{3}{2},3,\tfrac{7}{2}\right\},
\left\{\tfrac{3}{2},\tfrac{7}{2},6\right\},
\left\{\tfrac{5}{2},\tfrac{7}{2},4\right\},
\left\{3,\tfrac{7}{2},4\right\}$\\ &$
\left\{3,\tfrac{7}{2},5\right\},
\left\{\tfrac{7}{2},4,6\right\},
\left\{\tfrac{7}{2},4,10\right\},
\left\{\tfrac{7}{2},6,12\right\}$ \\ \hline
         
    $\tfrac{18}{5}$ & $\left\{\tfrac{6}{5},\tfrac{12}{5},\tfrac{18}{5}\right\},
\left\{\tfrac{6}{5},\tfrac{18}{5},6\right\},
\left\{\tfrac{18}{5},6,8\right\},
\left\{\tfrac{18}{5},6,12\right\}$ \\ \hline
         
    $\tfrac{9}{2}$ &$\left\{\tfrac{3}{2},3,\tfrac{9}{2}\right\},
\left\{3,\tfrac{9}{2},5\right\},
\left\{3,\tfrac{9}{2},6\right\},
\left\{\tfrac{9}{2},6,8\right\},
\left\{\tfrac{9}{2},6,12\right\}$ \\ \hline
         
    $\tfrac{14}{3}$& $\left\{\tfrac{4}{3},\tfrac{8}{3},\tfrac{14}{3}\right\},
\left\{\tfrac{4}{3},4,\tfrac{14}{3}\right\},
\left\{\tfrac{8}{3},\tfrac{14}{3},6\right\},
\left\{4,\tfrac{14}{3},6\right\},
\left\{4,\tfrac{14}{3},10\right\},
\left\{\tfrac{14}{3},6,12\right\}$\\ \hline
         
    $7$ & $\left\{3,4,7\right\},
\left\{3,5,7\right\},
\left\{4,6,7\right\},
\left\{4,7,10\right\},
\left\{6,7,12\right\}$ \\ \hline
         
    $9$& $\left\{3,5,9\right\},
\left\{3,6,9\right\},
\left\{6,8,9\right\},
\left\{6,9,12\right\}$\\ \hline
         
    $14$ & $\left\{4,6,14\right\},
\left\{4,10,14\right\},
\left\{6,12,14\right\}$ \\ \hline
         
    $18$ & $\left\{6,8,18\right\},
\left\{6,12,18\right\}$ \\ \hline\hline
\end{tabular}
\caption{The genuinely rank-$3$ sets of scaling dimensions with $\tfrac{7}{4}\leq \Delta_{\mathrm{new}}\leq 18$. 
The two triples in blue are repeated from table \ref{tab:rank3_part1} since $7/5$ is also a genuinely rank-3 dimension.}
\label{tab:rank3_part2}
\end{sidewaystable}

\begin{sidewaystable}
\centering
\begin{tabular}{cc}\hline\hline
         $\Delta_{\mathrm{new}}$&  $\{\Delta_1,\Delta_2,\Delta_3,\Delta_4\}$ \\ \hline
         
        $\tfrac{30}{29}$ & --- \\ \hline
        
        $\tfrac{24}{23}$ & $\left\{\tfrac{24}{23},6,8,12\right\},
\left\{\tfrac{24}{23},6,12,18\right\}$ \\ \hline
        
        {$\tfrac{20}{19}$} & $\left\{\tfrac{20}{19},4,8,10\right\},
\left\{\tfrac{20}{19},4,8,12\right\},
\left\{\tfrac{20}{19},4,10,14\right\}$ \\ \hline
        
        \multirow{5}{*}{$\tfrac{16}{15}$} &$\left\{\tfrac{16}{15},\tfrac{6}{5},\tfrac{4}{3},\tfrac{8}{5}\right\},
\left\{\tfrac{16}{15},\tfrac{6}{5},\tfrac{4}{3},\tfrac{8}{3}\right\},
\left\{\tfrac{16}{15},\tfrac{6}{5},\tfrac{8}{5},\tfrac{12}{5}\right\},
\left\{\tfrac{16}{15},\tfrac{6}{5},\tfrac{8}{5},\tfrac{14}{5}\right\},
\left\{\tfrac{16}{15},\tfrac{6}{5},\tfrac{8}{5},6\right\},
\left\{\tfrac{16}{15},\tfrac{6}{5},\tfrac{12}{5},\tfrac{18}{5}\right\},
\left\{\tfrac{16}{15},\tfrac{6}{5},\tfrac{8}{3},6\right\},
\left\{\tfrac{16}{15},\tfrac{6}{5},\tfrac{18}{5},6\right\}$\\ &$
\left\{\tfrac{16}{15},\tfrac{6}{5},6,8\right\},
\left\{\tfrac{16}{15},\tfrac{6}{5},8,12\right\},
\left\{\tfrac{16}{15},\tfrac{4}{3},\tfrac{8}{5},4\right\},
\left\{\tfrac{16}{15},\tfrac{4}{3},\tfrac{8}{3},\tfrac{10}{3}\right\},
\left\{\tfrac{16}{15},\tfrac{4}{3},\tfrac{8}{3},4\right\},
\left\{\tfrac{16}{15},\tfrac{4}{3},4,\tfrac{14}{3}\right\},
\left\{\tfrac{16}{15},\tfrac{4}{3},4,8\right\}$\\ &$
\left\{\tfrac{16}{15},\tfrac{4}{3},4,10\right\},
\left\{\tfrac{16}{15},\tfrac{8}{5},\tfrac{14}{5},4\right\},
\left\{\tfrac{16}{15},\tfrac{8}{5},4,6\right\},
\left\{\tfrac{16}{15},\tfrac{12}{5},4,8\right\},
\left\{\tfrac{16}{15},\tfrac{8}{3},4,6\right\},
\left\{\tfrac{16}{15},\tfrac{14}{5},4,10\right\}$\\ &$
\left\{\tfrac{16}{15},\tfrac{10}{3},6,8\right\},
\left\{\tfrac{16}{15},\tfrac{10}{3},8,12\right\},
\left\{\tfrac{16}{15},4,\tfrac{14}{3},6\right\},
\left\{\tfrac{16}{15},4,6,8\right\},
\left\{\tfrac{16}{15},4,6,10\right\},
\left\{\tfrac{16}{15},4,8,12\right\}$ \\ \hline

 $\tfrac{15}{14}$ & $\left\{\tfrac{15}{14},\tfrac{8}{7},8,12\right\},
\left\{\tfrac{15}{14},\tfrac{3}{2},3,\tfrac{9}{2}\right\},
\left\{\tfrac{15}{14},\tfrac{3}{2},3,5\right\},
\left\{\tfrac{15}{14},3,5,9\right\},
\left\{\tfrac{15}{14},3,8,12\right\}$ \\ \hline

 $\tfrac{15}{13}$ & $\left\{\tfrac{15}{13},3,5,9\right\},
\left\{\tfrac{15}{13},3,8,12\right\}$ \\ \hline 

$\tfrac{20}{17}$ & $\left\{\tfrac{20}{17},4,8,10\right\},
\left\{\tfrac{20}{17},4,8,12\right\},
\left\{\tfrac{20}{17},4,10,14\right\}$ \\  \hline 

$\tfrac{16}{13}$ & $\left\{\tfrac{14}{13},\tfrac{16}{13},4,10\right\},
\left\{\tfrac{16}{13},4,6,8\right\},
\left\{\tfrac{16}{13},4,6,10\right\},
\left\{\tfrac{16}{13},4,8,12\right\}$ \\ \hline

$\tfrac{24}{19}$ & $\left\{\tfrac{24}{19},6,8,12\right\},
\left\{\tfrac{24}{19},6,12,18\right\}$ \\ \hline

$\tfrac{30}{23}$ & --- \\ \hline

$\tfrac{15}{11}$ & $\left\{\tfrac{15}{11},3,5,9\right\},
\left\{\tfrac{15}{11},3,8,12\right\}$ \\ \hline

$\tfrac{24}{17}$ & $\left\{\tfrac{18}{17},\tfrac{24}{17},6,12\right\},
\left\{\tfrac{24}{17},6,8,12\right\},
\left\{\tfrac{24}{17},6,12,18\right\}$ \\ \hline

\multirow{2}{*}{$\tfrac{16}{11}$} & $\left\{\tfrac{12}{11},\tfrac{14}{11},\tfrac{16}{11},\tfrac{18}{11}\right\},
\left\{\tfrac{12}{11},\tfrac{14}{11},\tfrac{16}{11},6\right\},
\left\{\tfrac{12}{11},\tfrac{16}{11},\tfrac{18}{11},\tfrac{24}{11}\right\},
\left\{\tfrac{12}{11},\tfrac{16}{11},\tfrac{24}{11},6\right\},
\left\{\tfrac{12}{11},\tfrac{16}{11},6,8\right\},
\left\{\tfrac{12}{11},\tfrac{16}{11},8,12\right\}$\\ &$
\left\{\tfrac{14}{11},\tfrac{16}{11},4,6\right\},
\left\{\tfrac{16}{11},4,6,8\right\},
\left\{\tfrac{16}{11},4,6,10\right\},
\left\{\tfrac{16}{11},4,8,12\right\}$ \\ \hline

$\tfrac{20}{13}$ & $\left\{\tfrac{20}{13},4,8,10\right\},
\left\{\tfrac{20}{13},4,8,12\right\},
\left\{\tfrac{20}{13},4,10,14\right\}$ \\ \hline

$\tfrac{30}{19}$ & --- \\

\hline\hline
\end{tabular}
\caption{The genuinely rank-$4$ sets of scaling dimensions with $\tfrac{30}{29}\leq \Delta_{\mathrm{new}}\leq \tfrac{30}{19}$.}
\label{tab:rank4_part1}
\end{sidewaystable}

\begin{sidewaystable}
\centering
\begin{tabular}{cc}\hline\hline
         $\Delta_{\mathrm{new}}$&  $\{\Delta_1,\Delta_2,\Delta_3,\Delta_4\}$ \\ \hline

         $\tfrac{30}{17}$ & --- \\ \hline
         
         \multirow{4}{*}{$\tfrac{16}{9}$} & $\left\{\tfrac{10}{9},\tfrac{4}{3},\tfrac{14}{9},\tfrac{16}{9}\right\},
\left\{\tfrac{10}{9},\tfrac{4}{3},\tfrac{16}{9},\tfrac{20}{9}\right\},
\left\{\tfrac{10}{9},\tfrac{4}{3},\tfrac{16}{9},\tfrac{10}{3}\right\},
\left\{\tfrac{10}{9},\tfrac{4}{3},\tfrac{16}{9},4\right\},
\left\{\tfrac{10}{9},\tfrac{14}{9},\tfrac{16}{9},6\right\},
\left\{\tfrac{10}{9},\tfrac{16}{9},4,6\right\}$\\ &$
\left\{\tfrac{4}{3},\tfrac{14}{9},\tfrac{16}{9},\tfrac{8}{3}\right\},
\left\{\tfrac{4}{3},\tfrac{16}{9},\tfrac{8}{3},\tfrac{10}{3}\right\},
\left\{\tfrac{4}{3},\tfrac{16}{9},\tfrac{8}{3},4\right\},
\left\{\tfrac{4}{3},\tfrac{16}{9},4,\tfrac{14}{3}\right\},
\left\{\tfrac{4}{3},\tfrac{16}{9},4,8\right\},
\left\{\tfrac{4}{3},\tfrac{16}{9},4,10\right\},
\left\{\tfrac{14}{9},\tfrac{16}{9},\tfrac{8}{3},6\right\}$\\ &$
\left\{\tfrac{16}{9},\tfrac{20}{9},8,12\right\},
\left\{\tfrac{16}{9},\tfrac{8}{3},4,6\right\},
\left\{\tfrac{16}{9},\tfrac{10}{3},6,8\right\},
\left\{\tfrac{16}{9},\tfrac{10}{3},8,12\right\},
\left\{\tfrac{16}{9},4,\tfrac{14}{3},6\right\}$\\ &$
\left\{\tfrac{16}{9},4,6,8\right\},
\left\{\tfrac{16}{9},4,6,10\right\},
\left\{\tfrac{16}{9},4,8,12\right\}$ \\ \hline

$\tfrac{20}{11}$ & $\left\{\tfrac{12}{11},\tfrac{20}{11},4,8\right\},
\left\{\tfrac{14}{11},\tfrac{20}{11},4,10\right\},
\left\{\tfrac{20}{11},4,8,10\right\},
\left\{\tfrac{20}{11},4,8,12\right\},
\left\{\tfrac{20}{11},4,10,14\right\}$ \\ \hline

$\tfrac{24}{13}$ & $\left\{\tfrac{14}{13},\tfrac{18}{13},\tfrac{24}{13},6\right\},
\left\{\tfrac{14}{13},\tfrac{24}{13},6,8\right\},
\left\{\tfrac{18}{13},\tfrac{24}{13},6,12\right\},
\left\{\tfrac{24}{13},\tfrac{30}{13},8,12\right\},
\left\{\tfrac{24}{13},6,8,12\right\},
\left\{\tfrac{24}{13},6,12,18\right\}$ \\ \hline

$\tfrac{15}{8}$ & $\left\{\tfrac{9}{8},\tfrac{5}{4},\tfrac{3}{2},\tfrac{15}{8}\right\},
\left\{\tfrac{5}{4},\tfrac{3}{2},\tfrac{15}{8},3\right\},
\left\{\tfrac{3}{2},\tfrac{15}{8},3,\tfrac{9}{2}\right\},
\left\{\tfrac{3}{2},\tfrac{15}{8},3,5\right\},
\left\{\tfrac{15}{8},3,5,9\right\},
\left\{\tfrac{15}{8},3,8,12\right\}$\\ \hline

$\tfrac{15}{7}$ & $\left\{\tfrac{8}{7},\tfrac{15}{7},8,12\right\},
\left\{\tfrac{15}{7},3,5,9\right\},
\left\{\tfrac{15}{7},3,8,12\right\}$ \\ \hline

$\tfrac{24}{11}$ & $\blue{\left\{\tfrac{12}{11},\tfrac{16}{11},\tfrac{18}{11},\tfrac{24}{11}\right\}},
\blue{\left\{\tfrac{12}{11},\tfrac{16}{11},\tfrac{24}{11},6\right\}},
\left\{\tfrac{12}{11},\tfrac{18}{11},\tfrac{24}{11},\tfrac{30}{11}\right\},
\left\{\tfrac{12}{11},\tfrac{24}{11},6,8\right\},
\left\{\tfrac{24}{11},6,8,12\right\},
\left\{\tfrac{24}{11},6,12,18\right\}$ \\ \hline

\multirow{2}{*}{$\tfrac{20}{9}$} & $\blue{\left\{\tfrac{10}{9},\tfrac{4}{3},\tfrac{16}{9},\tfrac{20}{9}\right\}},
\left\{\tfrac{10}{9},\tfrac{4}{3},\tfrac{20}{9},4\right\},
\left\{\tfrac{10}{9},\tfrac{20}{9},4,8\right\},
\left\{\tfrac{4}{3},\tfrac{20}{9},\tfrac{10}{3},4\right\},
\left\{\tfrac{4}{3},\tfrac{20}{9},4,10\right\}$\\ &$
\blue{\left\{\tfrac{16}{9},\tfrac{20}{9},8,12\right\}},
\left\{\tfrac{20}{9},\tfrac{10}{3},4,8\right\},
\left\{\tfrac{20}{9},4,8,10\right\},
\left\{\tfrac{20}{9},4,8,12\right\},
\left\{\tfrac{20}{9},4,10,14\right\}$ \\ \hline

\multirow{2}{*}{$\tfrac{16}{7}$} & $\left\{\tfrac{8}{7},\tfrac{10}{7},\tfrac{12}{7},\tfrac{16}{7}\right\},
\left\{\tfrac{8}{7},\tfrac{10}{7},\tfrac{16}{7},4\right\},
\left\{\tfrac{8}{7},\tfrac{12}{7},\tfrac{16}{7},6\right\},
\left\{\tfrac{8}{7},\tfrac{16}{7},4,6\right\},
\left\{\tfrac{10}{7},\tfrac{16}{7},4,8\right\},
\left\{\tfrac{10}{7},\tfrac{16}{7},4,10\right\}$\\ &$
\left\{\tfrac{12}{7},\tfrac{16}{7},6,8\right\},
\left\{\tfrac{12}{7},\tfrac{16}{7},8,12\right\},
\left\{\tfrac{16}{7},\tfrac{18}{7},6,8\right\},
\left\{\tfrac{16}{7},\tfrac{20}{7},4,10\right\},
\left\{\tfrac{16}{7},4,6,8\right\},
\left\{\tfrac{16}{7},4,6,10\right\},
\left\{\tfrac{16}{7},4,8,12\right\}$ \\ \hline

$\tfrac{30}{13}$ & $\blue{\left\{\tfrac{24}{13},\tfrac{30}{13},8,12\right\}}$ \\ \hline

$\tfrac{30}{11}$ & $\blue{\left\{\tfrac{12}{11},\tfrac{18}{11},\tfrac{24}{11},\tfrac{30}{11}\right\}},
\left\{\tfrac{18}{11},\tfrac{30}{11},8,12\right\}$ \\ \hline

$\tfrac{20}{7}$ & $\left\{\tfrac{8}{7},\tfrac{10}{7},\tfrac{18}{7},\tfrac{20}{7}\right\},
\left\{\tfrac{8}{7},\tfrac{18}{7},\tfrac{20}{7},\tfrac{30}{7}\right\},
\left\{\tfrac{8}{7},\tfrac{20}{7},8,12\right\},
\left\{\tfrac{10}{7},\tfrac{20}{7},4,8\right\},
\blue{\left\{\tfrac{16}{7},\tfrac{20}{7},4,10\right\}},
\left\{\tfrac{20}{7},4,8,10\right\},
\left\{\tfrac{20}{7},4,8,12\right\},
\left\{\tfrac{20}{7},4,10,14\right\}$  \\
         
         \hline \hline
\end{tabular}
\caption{The genuinely rank-$4$ sets of scaling dimensions with $\tfrac{30}{17}\leq \Delta_{\mathrm{new}}\leq \tfrac{20}{7}$.}
\label{tab:rank4_part2}
\end{sidewaystable}

\begin{sidewaystable}
\centering
\begin{tabular}{cc}\hline\hline
         $\Delta_{\mathrm{new}}$&  $\{\Delta_1,\Delta_2,\Delta_3,\Delta_4\}$ \\ \hline

         \multirow{3}{*}{$\tfrac{16}{5}$} & $\left\{\tfrac{6}{5},\tfrac{8}{5},\tfrac{12}{5},\tfrac{16}{5}\right\},
\left\{\tfrac{6}{5},\tfrac{8}{5},\tfrac{14}{5},\tfrac{16}{5}\right\},
\left\{\tfrac{6}{5},\tfrac{8}{5},\tfrac{16}{5},6\right\},
\left\{\tfrac{6}{5},\tfrac{12}{5},\tfrac{16}{5},\tfrac{18}{5}\right\},
\left\{\tfrac{6}{5},\tfrac{12}{5},\tfrac{16}{5},\tfrac{24}{5}\right\},
\left\{\tfrac{6}{5},\tfrac{14}{5},\tfrac{16}{5},\tfrac{18}{5}\right\}$\\ &$
\left\{\tfrac{6}{5},\tfrac{16}{5},\tfrac{18}{5},6\right\},
\left\{\tfrac{6}{5},\tfrac{16}{5},6,8\right\},
\left\{\tfrac{6}{5},\tfrac{16}{5},8,12\right\},
\left\{\tfrac{8}{5},\tfrac{12}{5},\tfrac{16}{5},4\right\},
\left\{\tfrac{8}{5},\tfrac{14}{5},\tfrac{16}{5},4\right\},
\left\{\tfrac{8}{5},\tfrac{16}{5},4,6\right\},
\left\{\tfrac{12}{5},\tfrac{16}{5},4,8\right\}$\\ &$
\left\{\tfrac{12}{5},\tfrac{16}{5},4,10\right\},
\left\{\tfrac{14}{5},\tfrac{16}{5},4,8\right\},
\left\{\tfrac{14}{5},\tfrac{16}{5},4,10\right\},
\left\{\tfrac{16}{5},4,6,8\right\},
\left\{\tfrac{16}{5},4,6,10\right\},
\left\{\tfrac{16}{5},4,8,12\right\}$ \\ \hline

         $\tfrac{24}{7}$ & $\left\{\tfrac{8}{7},\tfrac{12}{7},\tfrac{18}{7},\tfrac{24}{7}\right\},
\left\{\tfrac{8}{7},\tfrac{12}{7},\tfrac{24}{7},6\right\},
\left\{\tfrac{8}{7},\tfrac{24}{7},6,12\right\},
\left\{\tfrac{12}{7},\tfrac{24}{7},6,8\right\},
\left\{\tfrac{18}{7},\tfrac{24}{7},8,12\right\},
\left\{\tfrac{24}{7},6,8,12\right\},
\left\{\tfrac{24}{7},6,12,18\right\}$ \\ \hline

$\tfrac{15}{4}$ & $\left\{\tfrac{5}{4},\tfrac{3}{2},3,\tfrac{15}{4}\right\},
\left\{\tfrac{3}{2},3,\tfrac{15}{4},\tfrac{9}{2}\right\},
\left\{\tfrac{3}{2},3,\tfrac{15}{4},5\right\},
\left\{3,\tfrac{15}{4},5,9\right\},
\left\{3,\tfrac{15}{4},8,12\right\}$ \\ \hline

$\tfrac{30}{7}$ & $\blue{\left\{\tfrac{8}{7},\tfrac{18}{7},\tfrac{20}{7},\tfrac{30}{7}\right\}},
\left\{\tfrac{8}{7},\tfrac{30}{7},8,12\right\}$ \\\hline

\multirow{2}{*}{$\tfrac{24}{5}$} & $\blue{\left\{\tfrac{6}{5},\tfrac{12}{5},\tfrac{16}{5},\tfrac{24}{5}\right\}},
\left\{\tfrac{6}{5},\tfrac{12}{5},\tfrac{18}{5},\tfrac{24}{5}\right\},
\left\{\tfrac{6}{5},\tfrac{24}{5},8,12\right\},
\left\{\tfrac{12}{5},\tfrac{18}{5},\tfrac{24}{5},6\right\}$\\ &$
\left\{\tfrac{12}{5},\tfrac{24}{5},6,8\right\},
\left\{\tfrac{18}{5},\tfrac{24}{5},6,12\right\},
\left\{\tfrac{24}{5},6,8,12\right\},
\left\{\tfrac{24}{5},6,12,18\right\}$ \\ \hline

\multirow{2}{*}{$\tfrac{16}{3}$} & $\left\{\tfrac{4}{3},\tfrac{8}{3},\tfrac{10}{3},\tfrac{16}{3}\right\},
\left\{\tfrac{4}{3},\tfrac{8}{3},4,\tfrac{16}{3}\right\},
\left\{\tfrac{4}{3},4,\tfrac{14}{3},\tfrac{16}{3}\right\},
\left\{\tfrac{4}{3},4,\tfrac{16}{3},8\right\},
\left\{\tfrac{4}{3},4,\tfrac{16}{3},10\right\},
\left\{\tfrac{8}{3},4,\tfrac{16}{3},6\right\},
\left\{\tfrac{10}{3},\tfrac{16}{3},8,12\right\}$\\ &$
\left\{4,\tfrac{14}{3},\tfrac{16}{3},6\right\},
\left\{4,\tfrac{16}{3},6,8\right\},
\left\{4,\tfrac{16}{3},6,10\right\},
\left\{4,\tfrac{16}{3},\tfrac{20}{3},8\right\},
\left\{4,\tfrac{16}{3},8,12\right\}$ \\ \hline

$\tfrac{20}{3}$ & $\left\{\tfrac{4}{3},\tfrac{10}{3},4,\tfrac{20}{3}\right\},
\left\{\tfrac{4}{3},4,\tfrac{20}{3},10\right\},
\left\{\tfrac{10}{3},4,\tfrac{20}{3},8\right\},
\blue{\left\{4,\tfrac{16}{3},\tfrac{20}{3},8\right\}},
\left\{4,\tfrac{20}{3},8,10\right\},
\left\{4,\tfrac{20}{3},8,12\right\},
\left\{4,\tfrac{20}{3},10,14\right\}$ \\ \hline

$\tfrac{15}{2}$ & $\left\{\tfrac{3}{2},3,\tfrac{9}{2},\tfrac{15}{2}\right\},
\left\{\tfrac{3}{2},3,5,\tfrac{15}{2}\right\},
\left\{3,5,\tfrac{15}{2},9\right\},
\left\{3,\tfrac{15}{2},8,12\right\}$ \\ \hline

$15$ & $\left\{3,5,9,15\right\},
\left\{3,8,12,15\right\}$ \\ \hline

$16$ & $\left\{4,6,8,16\right\},
\left\{4,6,10,16\right\},
\left\{4,8,12,16\right\}$ \\ \hline

$20$ & $\left\{4,8,10,20\right\},
\left\{4,8,12,20\right\},
\left\{4,10,14,20\right\},
\left\{4,12,14,20\right\},
\left\{8,12,20,24\right\},
\left\{12,14,18,20\right\}$ \\ \hline

$24$ & $\left\{6,8,12,24\right\},
\left\{6,12,18,24\right\},
\blue{\left\{8,12,20,24\right\}},
\left\{12,18,24,30\right\}$ \\ \hline

$30$ & $\blue{\left\{12,18,24,30\right\}}$ \\

\hline\hline
\end{tabular}
\caption{The genuinely rank-$4$ sets of scaling dimensions with $\tfrac{16}{5}\leq \Delta_{\mathrm{new}}\leq 30$.}
\label{tab:rank4_part3}
\end{sidewaystable}

\clearpage
\bibliographystyle{JHEP} 
\bibliography{References} 

\providecommand{\href}[2]{#2}\begingroup\raggedright\begin{thebibliography}{10}

\bibitem{Caorsi:2018zsq}
M.~Caorsi and S.~Cecotti, {\it {Geometric classification of 4d $\mathcal{N}=2$
  SCFTs}},  {\em JHEP} {\bf 07} (2018) 138,
  [\href{http://arxiv.org/abs/1801.04542}{{\tt arXiv:1801.04542}}].

\bibitem{Argyres:2018urp}
P.~C. Argyres and M.~Martone, {\it {Scaling dimensions of Coulomb branch
  operators of 4d N=2 superconformal field theories}},
  \href{http://arxiv.org/abs/1801.06554}{{\tt arXiv:1801.06554}}.

\bibitem{Kaidi:2022sng}
J.~Kaidi, M.~Martone, L.~Rastelli, and M.~Weaver, {\it {Needles in a haystack.
  An algorithmic approach to the classification of 4d $ \mathcal{N} $ = 2
  SCFTs}},  {\em JHEP} {\bf 03} (2022) 210,
  [\href{http://arxiv.org/abs/2202.06959}{{\tt arXiv:2202.06959}}].

\bibitem{Cecotti:2023ksl}
S.~Cecotti, {\it {Direct and Inverse Problems in Special Geometry}},
  \href{http://arxiv.org/abs/2312.02536}{{\tt arXiv:2312.02536}}.

\bibitem{Argyres:2022lah}
P.~C. Argyres and M.~Martone, {\it {The rank 2 classification problem I: scale
  invariant geometries}},  \href{http://arxiv.org/abs/2209.09248}{{\tt
  arXiv:2209.09248}}.

\bibitem{Martone:2021ixp}
M.~Martone, {\it {Testing our understanding of SCFTs: a catalogue of rank-2 $
  \mathcal{N} $ = 2 theories in four dimensions}},  {\em JHEP} {\bf 07} (2022)
  123, [\href{http://arxiv.org/abs/2102.02443}{{\tt arXiv:2102.02443}}].

\bibitem{Argyres:2015ffa}
P.~Argyres, M.~Lotito, Y.~L\"u, and M.~Martone, {\it {Geometric constraints on
  the space of $ \mathcal{N} $ = 2 SCFTs. Part I: physical constraints on
  relevant deformations}},  {\em JHEP} {\bf 02} (2018) 001,
  [\href{http://arxiv.org/abs/1505.04814}{{\tt arXiv:1505.04814}}].

\bibitem{Argyres:2015gha}
P.~C. Argyres, M.~Lotito, Y.~L\"u, and M.~Martone, {\it {Geometric constraints
  on the space of $ \mathcal{N} $ = 2 SCFTs. Part II: construction of special
  K\"ahler geometries and RG flows}},  {\em JHEP} {\bf 02} (2018) 002,
  [\href{http://arxiv.org/abs/1601.00011}{{\tt arXiv:1601.00011}}].

\bibitem{Argyres:2016xmc}
P.~Argyres, M.~Lotito, Y.~L\"u, and M.~Martone, {\it {Geometric constraints on
  the space of $ \mathcal{N}$ = 2 SCFTs. Part III: enhanced Coulomb branches
  and central charges}},  {\em JHEP} {\bf 02} (2018) 003,
  [\href{http://arxiv.org/abs/1609.04404}{{\tt arXiv:1609.04404}}].

\bibitem{Argyres:2020wmq}
P.~C. Argyres and M.~Martone, {\it {Towards a classification of rank r$
  \mathcal{N} $ = 2 SCFTs. Part II. Special Kahler stratification of the
  Coulomb branch}},  {\em JHEP} {\bf 12} (2020) 022,
  [\href{http://arxiv.org/abs/2007.00012}{{\tt arXiv:2007.00012}}].

\bibitem{Argyres:2022puv}
P.~C. Argyres and M.~Martone, {\it {The rank 2 classification problem II:
  mapping scale-invariant solutions to SCFTs}},
  \href{http://arxiv.org/abs/2209.09911}{{\tt arXiv:2209.09911}}.

\bibitem{Argyres:2022fwy}
P.~C. Argyres and M.~Martone, {\it {The rank-2 classification problem III:
  curves with additional automorphisms}},
  \href{http://arxiv.org/abs/2209.10555}{{\tt arXiv:2209.10555}}.

\bibitem{Alvarez-Gaume:1996ohl}
L.~Alvarez-Gaume and S.~F. Hassan, {\it {Introduction to S duality in N=2
  supersymmetric gauge theories: A Pedagogical review of the work of Seiberg
  and Witten}},  {\em Fortsch. Phys.} {\bf 45} (1997) 159--236,
  [\href{http://arxiv.org/abs/hep-th/9701069}{{\tt hep-th/9701069}}].

\bibitem{Lerche:1996xu}
W.~Lerche, {\it {Introduction to Seiberg-Witten theory and its stringy
  origin}},  {\em Nucl. Phys. B Proc. Suppl.} {\bf 55} (1997) 83--117,
  [\href{http://arxiv.org/abs/hep-th/9611190}{{\tt hep-th/9611190}}].

\bibitem{Freed:1997dp}
D.~S. Freed, {\it {Special Kahler manifolds}},  {\em Commun. Math. Phys.} {\bf
  203} (1999) 31--52, [\href{http://arxiv.org/abs/hep-th/9712042}{{\tt
  hep-th/9712042}}].

\bibitem{Martone:2020hvy}
M.~Martone, {\it {The constraining power of Coulomb Branch Geometry: lectures
  on Seiberg-Witten theory}},  in {\em {Young Researchers Integrability School
  and Workshop 2020}: {A modern primer for superconformal field theories}}, 6,
  2020.
\newblock \href{http://arxiv.org/abs/2006.14038}{{\tt arXiv:2006.14038}}.

\bibitem{Seiberg:1994aj}
N.~Seiberg and E.~Witten, {\it {Monopoles, duality and chiral symmetry breaking
  in N=2 supersymmetric QCD}},  {\em Nucl. Phys. B} {\bf 431} (1994) 484--550,
  [\href{http://arxiv.org/abs/hep-th/9408099}{{\tt hep-th/9408099}}].

\bibitem{Seiberg:1994rs}
N.~Seiberg and E.~Witten, {\it {Electric - magnetic duality, monopole
  condensation, and confinement in N=2 supersymmetric Yang-Mills theory}},
  {\em Nucl. Phys. B} {\bf 426} (1994) 19--52,
  [\href{http://arxiv.org/abs/hep-th/9407087}{{\tt hep-th/9407087}}]. [Erratum:
  Nucl.Phys.B 430, 485--486 (1994)].

\bibitem{Beem:2014zpa}
C.~Beem, M.~Lemos, P.~Liendo, L.~Rastelli, and B.~C. van Rees, {\it {The $
  \mathcal{N}=2 $ superconformal bootstrap}},  {\em JHEP} {\bf 03} (2016) 183,
  [\href{http://arxiv.org/abs/1412.7541}{{\tt arXiv:1412.7541}}].

\bibitem{Argyres:2017tmj}
P.~C. Argyres, Y.~L\"u, and M.~Martone, {\it {Seiberg-Witten geometries for
  Coulomb branch chiral rings which are not freely generated}},  {\em JHEP}
  {\bf 06} (2017) 144, [\href{http://arxiv.org/abs/1704.05110}{{\tt
  arXiv:1704.05110}}].

\bibitem{Freed:2012bs}
D.~S. Freed and C.~Teleman, {\it {Relative quantum field theory}},  {\em
  Commun. Math. Phys.} {\bf 326} (2014) 459--476,
  [\href{http://arxiv.org/abs/1212.1692}{{\tt arXiv:1212.1692}}].

\bibitem{Aharony:2013hda}
O.~Aharony, N.~Seiberg, and Y.~Tachikawa, {\it {Reading between the lines of
  four-dimensional gauge theories}},  {\em JHEP} {\bf 08} (2013) 115,
  [\href{http://arxiv.org/abs/1305.0318}{{\tt arXiv:1305.0318}}].

\bibitem{Gaiotto:2010be}
D.~Gaiotto, G.~W. Moore, and A.~Neitzke, {\it {Framed BPS States}},  {\em Adv.
  Theor. Math. Phys.} {\bf 17} (2013), no.~2 241--397,
  [\href{http://arxiv.org/abs/1006.0146}{{\tt arXiv:1006.0146}}].

\bibitem{DelZotto:2022ras}
M.~Del~Zotto and I.~Garc\'\i{}a~Etxebarria, {\it {Global structures from the
  infrared}},  {\em JHEP} {\bf 11} (2023) 058,
  [\href{http://arxiv.org/abs/2204.06495}{{\tt arXiv:2204.06495}}].

\bibitem{Argyres:2022kon}
P.~C. Argyres, M.~Martone, and M.~Ray, {\it {Dirac pairings, one-form
  symmetries and Seiberg-Witten geometries}},  {\em JHEP} {\bf 09} (2022) 020,
  [\href{http://arxiv.org/abs/2204.09682}{{\tt arXiv:2204.09682}}].

\bibitem{Closset:2023pmc}
C.~Closset and H.~Magureanu, {\it {Reading between the rational sections:
  Global structures of 4d $\mathcal{N}=2$ KK theories}},  {\em SciPost Phys.}
  {\bf 16} (2024) 137, [\href{http://arxiv.org/abs/2308.10225}{{\tt
  arXiv:2308.10225}}].

\bibitem{Argyres:2018zay}
P.~C. Argyres, C.~Long, and M.~Martone, {\it {The Singularity Structure of
  Scale-Invariant Rank-2 Coulomb Branches}},  {\em JHEP} {\bf 05} (2018) 086,
  [\href{http://arxiv.org/abs/1801.01122}{{\tt arXiv:1801.01122}}].

\bibitem{Donagi:1995am}
R.~Donagi and E.~Markman, {\it {Spectral curves, algebraically completely
  integrable Hamiltonian systems, and moduli of bundles}},
  \href{http://arxiv.org/abs/alg-geom/9507017}{{\tt alg-geom/9507017}}.

\bibitem{Donagi:1995cf}
R.~Donagi and E.~Witten, {\it {Supersymmetric Yang-Mills theory and integrable
  systems}},  {\em Nucl. Phys. B} {\bf 460} (1996) 299--334,
  [\href{http://arxiv.org/abs/hep-th/9510101}{{\tt hep-th/9510101}}].

\bibitem{lange2013complex}
H.~Lange and C.~Birkenhake, {\em Complex Abelian Varieties}.
\newblock Grundlehren der mathematischen Wissenschaften. Springer Berlin
  Heidelberg, 2013.

\bibitem{Argyres:2024hdn}
P.~C. Argyres, R.~Moscrop, S.~Thakur, and M.~Weaver, {\it {Completely
  (iso-)split scale-invariant Coulomb branch geometries are isotrivial}},
  \href{http://arxiv.org/abs/2405.19395}{{\tt arXiv:2405.19395}}.

\bibitem{Witten:1978mh}
E.~Witten and D.~I. Olive, {\it {Supersymmetry Algebras That Include
  Topological Charges}},  {\em Phys. Lett. B} {\bf 78} (1978) 97--101.

\bibitem{Martone:2020nsy}
M.~Martone, {\it {Towards the classification of rank-r$ \mathcal{N} $ = 2
  SCFTs. Part I. Twisted partition function and central charge formulae}},
  {\em JHEP} {\bf 12} (2020) 021, [\href{http://arxiv.org/abs/2006.16255}{{\tt
  arXiv:2006.16255}}].

\bibitem{Cecotti:2021ouq}
S.~Cecotti, M.~Del~Zotto, M.~Martone, and R.~Moscrop, {\it {The Characteristic
  Dimension of Four-Dimensional ${\mathcal {N}}$~=~2 SCFTs}},  {\em Commun.
  Math. Phys.} {\bf 400} (2023), no.~1 519--540,
  [\href{http://arxiv.org/abs/2108.10884}{{\tt arXiv:2108.10884}}].

\bibitem{HO2007}
J.-M. Hwang and K.~Oguiso, {\it {Characteristic foliation on the discriminant
  hypersurface of a holomorphic Lagrangian fibration}},  {\em Amer. J. Math.}
  {\bf 131} (2009) 981--1007, [\href{http://arxiv.org/abs/0710.2376}{{\tt
  arXiv:0710.2376}}].

\bibitem{HO2009}
J.-M. Hwang and K.~Oguiso, {\it {Multiple fibers of holomorphic Lagrangian
  fibrations}},  {\em Commun. Contemp. Math.} {\bf 13} (2011) 309--329,
  [\href{http://arxiv.org/abs/0907.4869}{{\tt arXiv:0907.4869}}].

\bibitem{Eguchi:1996ds}
T.~Eguchi and K.~Hori, {\it {N=2 superconformal field theories in
  four-dimensions and A-D-E classification}},  in {\em {Conference on the
  Mathematical Beauty of Physics (In Memory of C. Itzykson)}}, pp.~67--82, 7,
  1996.
\newblock \href{http://arxiv.org/abs/hep-th/9607125}{{\tt hep-th/9607125}}.

\bibitem{Shapere:1999xr}
A.~D. Shapere and C.~Vafa, {\it {BPS structure of Argyres-Douglas
  superconformal theories}},  \href{http://arxiv.org/abs/hep-th/9910182}{{\tt
  hep-th/9910182}}.

\bibitem{Giacomelli:2024dbd}
S.~Giacomelli, R.~Savelli, and G.~Zoccarato, {\it {$\mathcal{N} = 2$
  Orbi-S-Folds}},  \href{http://arxiv.org/abs/2405.00101}{{\tt
  arXiv:2405.00101}}.

\bibitem{Chacaltana:2010ks}
O.~Chacaltana and J.~Distler, {\it {Tinkertoys for Gaiotto Duality}},  {\em
  JHEP} {\bf 11} (2010) 099, [\href{http://arxiv.org/abs/1008.5203}{{\tt
  arXiv:1008.5203}}].

\bibitem{Perlmutter:2024noo}
E.~Perlmutter, {\it {A Rigorous Holographic Bound on AdS Scale Separation}},
  \href{http://arxiv.org/abs/2402.19358}{{\tt arXiv:2402.19358}}.

\end{thebibliography}\endgroup

\end{document}